 \documentclass[prd,aps,a4paper,twocolumn,nofootinbib]{revtex4-2}

\def\rtx@apsprd{%
\class@info{APS journal PRD selected}%
}%

\usepackage{graphicx,psfrag}
\usepackage{mathrsfs}
\usepackage{amsmath,amsfonts,amssymb}
\usepackage{multirow}
\usepackage{comment}
\usepackage{bm}
\usepackage{graphicx}
\usepackage{psfrag}
\usepackage{epstopdf}
\usepackage{color}
\usepackage{mathrsfs}
\usepackage{comment}
\usepackage{multirow}
\usepackage{hyperref}
\usepackage{diagbox} 
\usepackage{amsmath,amssymb,amsthm,wasysym}
\usepackage{capt-of}
\usepackage[utf8]{inputenc}
\usepackage[normalem]{ulem}

\def\mbar{\overline{m}}
\def\hr{{\hat{r}}}
\def\ha{{\hat{a}}}

\def\rmd{{\rm d}}

\newcommand{\abs}[1]{\left|#1 \right|}

\DeclareMathOperator{\sgn}{sgn}

\newcommand{\be}{\begin{equation}}
\newcommand{\ee}{\end{equation}}

\newcommand{\RN}[1]{%
  \textup{\uppercase\expandafter{\romannumeral#1}}%
}

\definecolor{gray}{rgb}{0.5,0.5,0.5}
\definecolor{cyan}{rgb}{0,0.9,0.9}
\definecolor{orange}{rgb}{0.9,0.5,0}
\definecolor{magenta}{rgb}{1,0,1}
\definecolor{purple}{rgb}{0.8,0.4,0.8}
\definecolor{darkgreen}{rgb}{0,.8,0}
\definecolor{turquoise}{rgb}{0.25,0.88,0.82}

\begin{document}

\interfootnotelinepenalty=10000
\raggedbottom

\title{Spinning test body orbiting around a Kerr black hole: Eccentric equatorial orbits and their asymptotic gravitational-wave fluxes
}

\author{Viktor Skoup\'y$^{1,\,2}$\thanks{vskoupy@gmail.com}}
\author{Georgios Lukes-Gerakopoulos$^1$\thanks{gglukes@gmail.com}}
\affiliation{${}^1$Astronomical Institute of the Czech Academy of Sciences, Bo\v{c}n\'{i} II 1401/1a, CZ-141 00 Prague, Czech Republic}   
\affiliation{${}^2$Institute of Theoretical Physics, Faculty of Mathematics and Physics, Charles University, 18000 Prague, Czech Republic}

\begin{abstract}
We use the frequency and time domain Teukolsky formalism to calculate gravitational-wave fluxes from a spinning body on a bound eccentric equatorial orbit around a Kerr black hole. The spinning body is represented as a point particle following the pole-dipole approximation of the Mathisson-Papapetrou-Dixon equations. Reformulating these equations we are not only able to find the trajectory of a spinning particle in terms of its constants of motion, but also to provide a method to calculate the azimuthal and the radial frequency of this trajectory. Using these orbital quantities, we introduce the machinery to calculate through the frequency domain Teukolsky formalism the energy and the angular momentum fluxes at infinity, and at the horizon, along with the gravitational strain at infinity. We crosscheck the results obtained from the frequency domain approach with the results obtained from a time domain Teukolsky equation solver called Teukode. 
\end{abstract}


\maketitle

\section{Introduction}
\label{sec:intro}

An extreme mass ratio inspiral (EMRI) is one of the most promising events expected to be detected with future space-based gravitational wave (GW) detectors like Laser Interferometer Space Antenna (LISA) \citep{LISA}. An EMRI occurs when a stellar mass compact object such as a black hole (BH) or a neutron star (secondary object) is trapped in the vicinity of a supermassive black hole (SMBH) (primary object). Due to gravitational radiation reaction  the secondary is slowly spiralling into the primary while emitting GWs. From these GWs it is possible to extract information about the EMRI system such as the masses of the objects, their spins etc. On a more fundamental physics level, EMRIs detection are expected to allow us to probe  the strong gravity regime around a SMBH \citep{Babak:2017}. 

Currently in order to extract information from a GW signal, when it is detected by the terrestrial observatories, it has to be uncovered from a dominating noise background. To achieve this, matched filtering is employed, i.e. waveform templates for a wide range of parameters are matched with the detected time series. It is expected that we will have to use matched filtering for GW signal received by LISA as well, but not to uncover the signal from the noise; in LISA's case we will use them to disentangle overlapping GW signals from simultaneously detected sources. Because of this, accurate models of the GW waveform templates are planned to be produced for a wide range of parameters.

To model GWs from an EMRI, first the trajectory of the secondary object must be reproduced. The standard way to do this is to apply the two timescale approximation \citep{Hinderer:2008}. In an EMRI the mass ratio $q \equiv \mu/M$  lies between $10^{-7}$ and $10^{-4}$, where $\mu$ is the secondary mass and $M$ is the primary mass. The energy changes at rate $\dot{E}/E = \mathcal{O}(q)$ which is very small. The timescale of the inspiral is, thus, of the order $\mathcal{O}(q^{-1})$, i.e, $q^{-1}$ times larger than the orbital timescale. This allows us to break our analysis in two timescales, the fast orbital and the slow adiabatic dissipation in the constants of motion. In the fast one, the trajectory of the secondary over one orbital period is close to a trajectory calculated without a dissipation. The secondary is actually drifting between orbits characterized by a set of constants of motion. In this setup, the azimuthal coordinate of the inspiral can be expanded as $\phi=q^{-1}\phi^{(0)}(qt) + \phi^{(1)}(qt) + \mathcal{O}(q)$. The first term of the expansion is of \emph{adiabatic order} and includes the contribution from the time-averaged dissipative part of the first-order self-force. The second term, which is of the order of radians is called \emph{post-adiabatic} and contains contributions from the conservative part of the first-order self-force, oscillating part of the dissipative part of the first order self-force as well as the time-averaged dissipative part of the second-order self-force. The spin of the secondary contributes to the post-adiabatic term as is of the order of $\mathcal{O}(q)$ \cite{Mino:1997bx,Akcay20}. In particular, for the spin magnitude $S$ of a secondary compact object, like a Kerr BH or a neutron star, holds that  $S\lesssim \mu^2$, hence the dimensionless spin parameter defined as $\sigma \equiv S/(\mu M) \leq q$ is of the same order as the mass ratio \citep{Hartl:2002ig}. The phase $\phi$ is approximately proportional to the phase of the GW. Hence, to accurately model the GW fluxes, all the aforementioned terms must be taken into account. 

In this work, we deal with the contribution of the secondary spin to the post-adiabatic term, in the case of bounded equatorial orbits around a Kerr BH. The description of a spinning test body moving on a curved background was for the first time studied in \citep{Frenkel1926,Lanczos1930,Mathisson:1937zz}. In particular, Mathisson \citep{Mathisson:2010} managed to write the stress-energy tensor of an extended test body as a sum of multipolar moments. When the body is sufficiently small and compact, then it is sufficient to take into account only the mass (monopole) and the spin (dipole) leading to what is known as the \emph{pole-dipole approximation}, which essentially reduces the body to a spinning test particle. Later on Papapetrou  \citep{Papapetrou:1951pa,Corinaldesi:1951pb} was able to employ the conservation law of the stress energy tensor $\nabla_\mu T^{\mu\nu}=0$ to derive the equations of motion for a spinning particle. Finally, these equations were rewritten by Tulczyjew \citep{tulczyjew1959motion}, Dixon \citep{Dixon:1964,Dixon:1970zz,Dixon:1970zza,dixon1974dynamics} and Wald \citep{Wald:1972sz} bringing them to their modern form. MPD equations have been studied in several works, see e.g. \citep{Semerak:1999qc,Kyrian:2007zz,Bini:2000vv,Hartl:2002ig,Hartl:2003da}. Particularly, these equations simplify when the particle is confined into the equatorial plane of the Kerr spacetime \citep{Shibata:1998}. In this case, the motion can be determined by the following constants of motion: the energy $E$, the component of the total angular momentum parallel to the axis of the central BH $J_z$, the mass of the secondary $\mu$ and the magnitude of its spin $S$.

In the present work, we rederive the equations of motion for a spinning particle in the equatorial plane in a reduced form. This allows us to find analytical formulas for the constants of motion dependence on the eccentricity and the semi-latus rectum and to provide a method to numerically calculate the fundamental frequencies. These results are then used to calculate the GW fluxes. To achieve this, we employ the Teukolsky formalism and solve the GWs perturbatively. Namely, we solve the Teukolsky equation (TE) both in the frequency and in the time domain with a spinning-particle as a source. In the frequency domain, the formulas providing the energy and the angular momentum fluxes to infinity and to the horizon from a spinning particle following equatorial trajectories are novel. While, for the calculations in the time domain, we introduce a new approach to simulate the spinning source making the computations more efficient. Due to the GW flux balance law in an EMRI, these fluxes equal to the rate of change of the constants of motion of an inspiraling spinning particle \citep{Mino:2003,Akcay20}. Hence, once these fluxes are obtained, then the adiabatic term with the spinning-particle contribution to the post-adiabatic term can be reconstructed. 

This paper is organized as follows. Section~\ref{sec:dynamics} briefs the dynamics of a spinning particle moving in a curved spacetime. After covering the basics, the equations of motion of a spinning particle are rederived in a reduced form appropriate for eccentric equatorial orbits in a Kerr BH background. Subsequently, the constants of motion and the frequencies are calculated. Section~\ref{sec:GWfluxes} reviews the Teukolsky formalism calculating the GW fluxes both in the frequency and the time domain. Finally, the frequency domain results are compared with the time domain results. To make the main text more readable, we have concentrated in a list all the dimensionless quantities we use in Appendix~\ref{sec:quantities}, Appendix~\ref{sec:coefficientsAi} provides all the explicit formulas for the frequency domain fluxes, while in Appendix~\ref{sec:comaparisonDrasco} our frequency domain results for a non-spinning object are compared with the ones of \citep{Drasco:2005kz}. Finally, Appendix~\ref{sec:tables} provides tables from the frequency domain calculations aiming to serve as reference for future works.

Throughout this paper, we use geometrized units where the speed of light and the gravitational constant are $c=G=1$. The Riemann tensor is defined as $R^\mu{}_{\nu\kappa\lambda} = \Gamma^{\mu}{}_{\nu\lambda,\kappa} - \Gamma^{\mu}{}_{\nu\kappa,\lambda} + \Gamma^{\mu}{}_{\rho\kappa} \Gamma^{\rho}{}_{\nu\lambda} - \Gamma^{\mu}{}_{\rho\lambda} \Gamma^{\rho}{}_{\nu\kappa}$ where the comma denotes partial derivative $U_{\mu,\nu} = \partial_{\nu} U_{\mu}$. A covariant derivative is denoted by a semicolon $U_{\mu;\nu}=\nabla_\nu U_{\mu}$ and ${\rm D}U^\mu/\rmd \tau = U^\mu{}_{;\nu} \rmd x^\nu/\rmd \tau$. The signature of the metric is $(-,+,+,+)$. Symmetrization of indices is denoted by round brackets $\Phi_{(\mu\nu)} = (\Phi_{\mu\nu} + \Phi_{\nu\mu})/2$. For some quantities we prefer to use their dimensionless counterparts. They are denoted by a hat, e.g. energy $\hat{E}=E/\mu$, radial coordinate $\hr=r/M$ etc (see Appendix~\ref{sec:quantities}).

\section{A Pole-Dipole Particle moving on the equatorial plane of a Kerr Black hole}
\label{sec:dynamics}

The motion of a spinning test object in a curved background is governed by the Mathisson-Papapetrou-Dixon (MPD) equations \citep{Mathisson:1937zz,Papapetrou:1951pa,Dixon:1964} which read
\begin{align} 
 \label{eq:MPEQs}
 \frac{{\rm D}P^\mu}{\rmd \tau} &= 
 - \dfrac{1}{2} \; {R^\mu}_{\nu\rho\sigma} \; v^\nu \; S^{\rho\sigma} \; , \\ \nonumber
 \frac{{\rm D}S^{\mu\nu}}{\rmd \tau} & = 
 P^\mu v^\nu - P^\nu v^\mu \; ,
\end{align}
where $P^\mu$ is the four-momentum of the particle, $R^\mu{}_{\nu\rho\sigma}$ is the Riemann tensor of the background spacetime, $v^\mu = \rmd x^\mu/\rmd \tau$ is the four-velocity, $S^{\mu\nu}$ is the spin tensor of the particle and ${\rm D}/\rmd \tau = v^\mu \nabla_\mu$ is the covariant derivative along the worldline parametrized by the proper time $\tau$.

The stress-energy tensor $T^{\mu\nu}$ for a spinning particle with its trajectory parametrized by the coordinate time $t$ reads \citep{Faye:2006gx}
\begin{equation}
    T^{\mu\nu} = \frac{1}{\sqrt{-g}} \left( \frac{P^{(\mu} v^{\nu)}}{v^t} \delta^3 - \nabla_\alpha \left( \frac{S^{\alpha(\mu} v^{\nu)}}{v^t} \delta^3 \right) \right)\, ,
\end{equation}
where for Boyer-Lindquist (BL) coordinates $\delta^3 = \delta(r-r_p(t)) \delta(\theta-\theta_p(t)) \delta(\phi-\phi_p(t))$ is the delta function located at the particle position $(r_p(t), \theta_p(t), \phi_p(t))$ parametrized by coordinate time. Note that by using the conservation law $T^{\mu\nu}{}_{;\nu}=0$, it is possible to retrieve the MPD equations.

Actually, the MPD system of equations is underdetermined. The physical implication of the latter fact is that the center of the mass of the spinning object is not defined. To close the system of equations and to define the centre of the mass, a spin supplementary condition (SSC) in the form $S^{\mu\nu} V_\mu = 0$ has to be specified, where $V_\mu$ is a timelike vector field. In this work, we use the Tulczyjew-Dixon (TD) SSC \citep{tulczyjew1959motion,Dixon:1970zz} 
\begin{align}
 S^{\mu\nu} P_\mu = 0  \; .
\end{align}
Under the TD SSC, the rest mass of the particle with respect to the four-momentun
\begin{equation} \label{eq:p_norm}
    \mu^2 = -P^\mu P_\mu
\end{equation}
and the magnitude of the spin
\begin{align}
 \label{eq:spinMagnitude}
 S^2 = \frac{1}{2} S^{\mu\nu}S_{\mu\nu}
\end{align}
are conserved quantities (see e.g. \citep{Semerak:1999qc}). The conservation of the above quantities is independent of the spacetime background. The symmetries of the spacetime introduce for each Killing vector $\xi^\mu$ a specific quantity 
\begin{align}\label{eq:MPD_killsym}
 C=\xi^\mu P_\mu-\frac12 \xi_{\mu;\nu} S^{\mu\nu} \, ,
\end{align}
which is conserved upon the evolution of the MPD equations.

Instead of the spin tensor, it is sometimes more convenient to use the spin four-vector
\begin{align}
\label{eq:SpinVect}
 S_\mu = -\frac{1}{2} \epsilon_{\mu\nu\rho\sigma}
          \, u^\nu \, S^{\rho\sigma} \; ,
\end{align}
where $\epsilon_{\mu\nu\rho\sigma}$ is the Levi-Civita tensor and $u^\nu := P^\nu/\mu$ is the specific four-momentum. The inverse relation of this equation reads 
\begin{align}
   S^{\rho\sigma}=-\epsilon^{\rho\sigma\gamma\delta} S_{\gamma} u_\delta \; . 
   \label{eq:SpinTens}
\end{align}
After substituting Eq.~\eqref{eq:SpinTens} into Eq.~\eqref{eq:spinMagnitude}, we can derive the relation for the spin magnitude in terms of the spin four-vector 
\begin{align}
\label{eq:SpinCons}
 S^2=S^\mu S_\mu \; .
\end{align}
The spin four-vector is from the definition \eqref{eq:SpinVect} orthogonal to the four-momentum $P_\mu S^\mu=0$, while from Eq.~\eqref{eq:SpinTens} one sees it is orthogonal also to the spin tensor $S^{\mu\nu}S_\mu = 0$. Finally, from Eq.~\eqref{eq:v_p_TUL} it can be shown that it is orthogonal to the four-velocity $v_\mu S^\mu=0$ as well.

Since the MPD equations do not provide an evolution equation for the four-velocity, it is convenient that for the TD SSC exists an explicit relation of the four-velocity in terms of the four-momentum and the spin tensor \citep{Ehlers1977}. This relation reads 
\begin{align}
 \label{eq:v_p_TUL}
 v^\mu = \frac{\textsf{m}}{\mu} \left(
          u^\mu + 
          \frac{ 2 \; S^{\mu\nu} R_{\nu\rho\kappa\lambda} u^\rho S^{\kappa\lambda}}
          {4 \mu^2 + R_{\alpha\beta\gamma\delta} S^{\alpha\beta} S^{\gamma\delta} }
          \right)  \; ,
\end{align}
where $\mathsf{m} = -P_\mu v^\mu$ is the rest mass with respect to the four-velocity $v^\mu$. This mass  $\mathsf{m}$ is not conserved under the TD SSC, but it is used to conserve the normalization $v^\mu v_\mu = -1$ during the MPD evolution. This leads to \citep{Witzany:2019}
\begin{equation} \label{eq:mass_vmu}
    \textsf{m} = \frac{\mathcal{A}\mu^2}{\sqrt{\mathcal{A}^2 \mu^2 - \mathcal{B}S^2}} \; ,
\end{equation}
where
\begin{align}
    \mathcal{A} &= 4\mu^2 + R_{\alpha\beta\gamma\delta} S^{\alpha\beta} S^{\gamma\delta} \; , \\
    \mathcal{B} &= 4 h^{\kappa\eta} R_{\kappa\iota\lambda\mu} P^\iota S^{\lambda\mu} R_{\eta \nu \omega \pi} P^\nu S^{\omega\pi} \; , \\
    h^\kappa{}_\eta &= \frac{1}{S^2} S^{\kappa\rho} S_{\eta\rho} \; .
\end{align}

\subsection{The Kerr spacetime background}

Since our work deals with the motion of a spinning in the Kerr spacetime, let us briefly introduce this spacetime. The Kerr geometry in BL coordinates $(t,r,\theta,\phi)$ is described by the metric
\begin{multline}
    \rmd s^2 = g_{tt}~\rmd t^2+2~g_{t\phi}~\rmd t~\rmd \phi + g_{\phi\phi}~\rmd \phi^2 \\
       + g_{rr}~\rmd r^2+g_{\theta\theta}~\rmd \theta^2 \; , \label{eq:LinEl}
\end{multline}
where the metric coefficients are
 \begin{eqnarray}
   g_{tt} &=&-\left(1-\frac{2 M r}{\Sigma}\right) \; ,\nonumber\\ 
   g_{t\phi} &=& -\frac{2 a M r \sin^2{\theta}}{\Sigma} \; ,\nonumber\\
   g_{\phi\phi} &=& \frac{(\varpi^4-a^2\Delta \sin^2\theta) \sin^2{\theta}}{\Sigma} \; , \label{eq:KerrMetric}\\
   g_{rr} &=& \frac{\Sigma}{\Delta} \; ,\nonumber\\
   g_{\theta\theta} &=& \Sigma \nonumber
 \end{eqnarray} 
with
 \begin{eqnarray}
  \Sigma &=& r^2+ a^2 \cos^2{\theta} \; ,\nonumber\\
  \Delta &=& \varpi^2-2 M r \; ,\nonumber \\ 
  \varpi^2 &=& r^2+a^2 \; . \label{eq:Kerrfunc} 
 \end{eqnarray}

The Kerr spacetime is stationary and axisymmetric. This provides two Killing vector fields, the timelike one $\xi^\mu_{(t)}$ and the spacelike one $\xi^\mu_{(\phi)}$. Due to these Killing vector fields, Eq.~\eqref{eq:MPD_killsym} provides two constants of motion. In particular, thanks to the timelike field, the energy
 \begin{align}\label{eq:EnCons}
  E &= -P_t+\frac12g_{t\mu,\nu}S^{\mu\nu}
 \end{align}
is conserved, and thanks to the spacelike field, the component of the total angular momentum parallel to the rotational axis of Kerr ($z$ axis)
 \begin{align}\label{eq:JzCons}
  J_z &= P_\phi-\frac12g_{\phi\mu,\nu}S^{\mu\nu}
 \end{align}
is conserved. These two conserved quantities can be used to parametrize the spinning particles orbits as discussed in Sec.~\ref{sec:ConMot}.

\subsection{Equatorial orbits}

\label{sec:equatorial_dynamics}

We are interested in equatorial orbits, where $\theta = \pi/2$. To constrain the body to the equatorial plane, the $v^\theta$ component of the four-velocity must be always zero. The orthogonality of the spin four-vector and the four-velocity $v_\mu S^\mu=0$ implies that in order to achieve $v^\theta=0$ for arbitrary equatorial orbit  all the components of the spin four
vector should be zero except from $S^\theta$, i.e.,
\begin{align}\label{eq:SpinAli}
  S_\mu=S_\theta \delta_\mu^\theta \; .
\end{align}
The spin is, therefore, parallel to the $z$ axis. From the orthogonality of the spin four-vector and the four-momentum $P_\mu S^\mu=0$, it holds that $P^\theta = 0$. 

From Eqs.~\eqref{eq:SpinCons} and \eqref{eq:SpinAli} it can be shown that $S_\theta = -\sqrt{g_{\theta\theta}} S$ where the sign is chosen such that the spin magnitude is positive (negative) when the spin is parallel (antiparallel) to the $z$ axis. Then, from Eq.~\eqref{eq:SpinTens} the only nonzero components of the spin tensor are
 \begin{align}
  S^{tr} &= -S^{rt} = -S~u_\phi \sqrt{-\frac{g_{\theta\theta}}{g}} = - \frac{S~u_\phi}{r}  \; , \nonumber \\
  S^{t\phi} &= -S^{\phi t} = S~u_r \sqrt{-\frac{g_{\theta\theta}}{g}}=  \frac{S~u_r}{r} \; , \nonumber \\
  S^{r\phi} &= -S^{\phi r} = -S~u_t \sqrt{-\frac{g_{\theta\theta}}{g}}= -\frac{S~u_t}{r}  \; , \label{eq:Smunu}
 \end{align}
where $g$ is determinant of the metric. For Kerr spacetime on equatorial plane, it holds $\sqrt{-g_{\theta\theta}/g} = 1/r$.

Let us recheck the setup for equatorial orbits in a Kerr background. The total derivative with respect to proper time of the $\theta$ component of four-momentum can be expressed from Eq. \eqref{eq:MPEQs}
\begin{equation}
    \frac{\rmd P^\theta}{\rmd \tau} = - \dfrac{1}{2} \; {R^\theta}_{\nu\rho\sigma} \; v^\nu \; S^{\rho\sigma} - \Gamma^{\theta}{}_{\nu\rho} P^\nu v^\rho .
\end{equation}
The right-hand side (rhs) of this equation is equal to zero on the equatorial plane. Furthermore, Eq.~\eqref{eq:v_p_TUL} reduces on the equatorial plane to  $v^\theta = (\mathsf{m}/\mu^2) P^\theta$. This implies that when $v^\theta=0$ then $P_\theta$ remains zero as well. Thus, the particle stays on the equatorial plane by just demanding that $v^\theta=0$.

From Eqs.~\eqref{eq:EnCons},~\eqref{eq:JzCons} and \eqref{eq:Smunu}, $P_t$ and $P_\phi$ can be expressed as functions of $E$ and $J_z$. These expressions in dimensionless quantities read
\begin{align}
  \label{eq:specific_linmom_CEO_TUL}
  u_t &=\frac{-\hat{E}-\dfrac{\sigma}{\hat{r}^3}(\hat{a}\hat{E}-\hat{J}_z)}{1-\dfrac{\sigma^2}{\hat{r}^3}} \; , \nonumber \\
  u_\phi &= M \frac{\hat{J}_z-\dfrac{\sigma}{\hat{r}^3}\left[\left(-\hat{a}^2+\hat{r}^3 \right)
  \hat{E} + \hat{a} \hat{J}_z\right]}{1-\dfrac{\sigma^2}{\hat{r}^3}} \; .
 \end{align}

When we restrict the motion to the equatorial plane, it is possible to reproduce the equations of motion for the spinning particle from Eqs.~\eqref{eq:v_p_TUL} and \eqref{eq:p_norm}. In particular, we can express $u^r$ from the normalization~\eqref{eq:p_norm} as function of $E$ and $J_z$ and thanks to the fact that it holds
\begin{equation}
   2 S^{r\nu} R_{\nu\rho\kappa\lambda} u^\rho S^{\kappa\lambda} = \frac{12\mu^2 \hat\Delta \sigma^2 x^2}{\hr^3 \Sigma_\sigma^2} u^r
\end{equation}
we can write the equations of motion as
\begin{subequations} \label{eq:EOM_equatorial}
\begin{align}
    \Sigma_\sigma \Lambda_\sigma \frac{\rmd \hat{t}}{\rmd \hat{\tau}} &= \frac{\textsf{m}}{\mu} V^t(\hr) \; , \label{eq:EOM_t}\\
    \Sigma_\sigma \Lambda_\sigma \frac{\rmd \hr}{\rmd \hat{\tau}} &= \frac{\textsf{m}}{\mu} V^r(\hr) = \pm \frac{\textsf{m}}{\mu} \sqrt{R_\sigma(\hr)} \; , \label{eq:EOM_r}\\
    \Sigma_\sigma \Lambda_\sigma \frac{\rmd \phi}{\rmd \hat{\tau}} &= \frac{\textsf{m}}{\mu} V^\phi(\hr) \, , \label{eq:EOM_phi}
\end{align}
where
\begin{align}
    \Sigma_\sigma &= \hr^2 \left( 1 - \frac{\sigma^2}{\hr^3} \right) \; , \\
    \Lambda_\sigma &= 1- \frac{3\sigma^2 \hr x^2}{\Sigma_\sigma^3} \; , \\
    V^t &= \ha\left( 1+\frac{3\sigma^2}{\hr \Sigma_\sigma} \right) x + \frac{\varpi^2}{\Delta} P_\sigma \; , \\
    R_\sigma &= P_\sigma^2 - \hat{\Delta} \left( \frac{\Sigma_\sigma^2}{\hr^2} +x^2 \right) \; , \\
    V^\phi &= \left( 1+\frac{3\sigma^2}{\hr \Sigma_\sigma} \right) x + \frac{\ha}{\hat{\Delta}} P_\sigma \; , \\
    P_\sigma &= \Sigma_\sigma \hat{E} - \left( \ha + \frac{\sigma}{\hr} \right) x \; , \\
    x &= \hat{J}_z - (\ha+\sigma)\hat{E} \; .
\end{align}
\end{subequations}
The rest mass with respect to $v^\mu$ can be expressed from~\eqref{eq:mass_vmu} as
\begin{equation} \label{eq:m_over_mu}
    \frac{\mathsf{m}}{\mu} = \Lambda_\sigma \sqrt{\frac{1-\frac{\sigma^2}{\hr^3}}{-1+2\Lambda_\sigma - (2 - \Lambda_\sigma)\frac{\sigma^2}{\hr^3}}} \; .
\end{equation}
This expression is identical to Eq. (49) in \citep{Piovano:2020}. Equations~\eqref{eq:EOM_equatorial} are identical to the equations (2.19)--(2.21) in \citep{Shibata:1998} up to the parametrization with $\rmd \tilde{\tau}/\rmd\tau = \mathsf{m}/\mu$ where $\tilde{\tau}$ is the parametrization used in \citep{Shibata:1998}. By dividing Eqs.~\eqref{eq:EOM_r} and \eqref{eq:EOM_phi} we obtain Eq. (19) in \citep{Harms:2015ixa}. Hence, we have checked the validity of the above equations.

To simplify the equations of motion, it is useful to reparametrize Eqs.~\eqref{eq:EOM_equatorial} with a time parameter $\lambda$ which is similar to the Mino time \citep{Mino:2003}. Equations~\eqref{eq:EOM_equatorial} and \eqref{eq:m_over_mu} imply that the relation between $\hat{\tau}$ and $\lambda$ is
\begin{equation}
    \frac{\rmd \hat{\tau}}{\rmd \lambda} = \hr^2 \sqrt{\left( 1-\frac{\sigma^2}{\hr^3} \right)\left( -1+2\Lambda_\sigma - (2-\Lambda_\sigma) \frac{\sigma^2}{\hr^3} \right)} \; .
\end{equation}
Then it holds $\rmd \hat{x}^\mu/\rmd \lambda = V^\mu$ where $\hat{x}^\mu = (\hat{t},\hr,\theta,\phi)$ with $V^\theta = 0$. $V^\mu$ can be interpreted as dimensionless four-velocity with respect to $\lambda$.

\subsection{Constants of motion as orbital parameters} \label{sec:ConMot}

Let us see how we can use the constants of motion $E,~J_z$ to parametrize bounded equatorial orbits.
To do that we have to find first the roots of Eq.~\eqref{eq:EOM_r}, which will lead us to the  turning points of an equatorial eccentric orbit. The function $\hr^4 R_\sigma(\hr)$ is an eighth order polynomial, hence it has generally 8 roots. At least four of these roots are real as in the nonspinning case, while four additional roots, which come from the secondary spin's terms, can be complex or real. 
From these roots the two outermost ones $0<\hr_1\leq\hr_2$ are the candidates for being the turning points we are seeking. Obviously for these two roots it has to hold that
\begin{equation} \label{eq:zero_conditions}
    R_\sigma(\hr_1) = 0 \; , \qquad R_\sigma(\hr_2) = 0\; .
\end{equation}
To have a bound equatorial orbit between these two roots, Eq.~\eqref{eq:EOM_r} implies that $R_\sigma(\hr)>0$ for $\hr_1 < \hr < \hr_2$. The latter can be true only if for the derivative of $R_\sigma(\hr)$ with respect to $\hr$ it holds that 
\begin{equation} \label{eq:dRdr_conditions}
    R'_\sigma(\hr_1) \geq 0 \; , \qquad R'_\sigma(\hr_2) < 0 \; .
\end{equation}
When the conditions~\eqref{eq:zero_conditions},~\eqref{eq:dRdr_conditions} are satisfied, then $\hr_1$ is the pericenter and $\hr_2$ is the apocenter of an equatorial eccentric orbit, and it also holds that $\hat{E}^2<1$.

Having found the turning points of an equatorial eccentric orbit, we can parametrize each eccentric equatorial orbit by its semi-latus rectum $p$ and its eccentricity $e$, which relate to the turning points as follows
\begin{equation}
    \hr_1 = \frac{p}{1+e} \; , \qquad \hr_2 = \frac{p}{1-e} \; .
\end{equation}
The inverse relations read
\begin{equation}
    p = \frac{2 \hr_1 \hr_2}{\hr_1 + \hr_2} \; , \qquad e = \frac{\hr_2 - \hr_1}{\hr_1 + \hr_2} \; .
\end{equation}

Equation~\eqref{eq:zero_conditions} can be written as two quadratic equations in terms of $\hat{E}$ and $\hat{J}_z$. Using the same method as in Appendix B of \citep{Schmidt:2002} we can rearrange the formulas~\eqref{eq:zero_conditions} for energy and angular momentum to arrive at
\begin{align}\label{eq:QuadConst}
    f_i \hat{E}^2 - 2 g_i \hat{E} \hat{J}_z - h_i \hat{J}_z^2 - d_i &= 0\qquad i=1,2
\end{align}
where $f_1 = f(\hr_1)$, $f_2 = f(\hr_2)$ etc. and
\begin{align}
f(\hr) =& \ha^2 (\hr+2) \hr + \hr^4 + \nonumber\\  &+ \sigma \left(\frac{\ha^2 \sigma}{\hr^2}+\frac{2 \ha^2 (\ha + \sigma)}{\hr} + 6 \ha \hr-(\hr-2) \hr \sigma \right) \nonumber\\
g(\hr) =& 2 \ha \hr + \sigma \left(\frac{\ha \sigma}{\hr^2} + \frac{\ha (2 \ha + \sigma)}{\hr}-(\hr-3) \hr\right) \\
h(\hr) =& \hat{\Delta} - \left(\ha+\frac{\sigma}{\hr}\right)^2 \nonumber\\
d(\hr) =& \frac{\hat{\Delta} (\hr^3 - \sigma^2)^2}{\hr^4} \nonumber
\end{align}
These functions for $\sigma = 0$ are identical to the functions (B.6) -- (B.9) in \citep{Schmidt:2002} with $z_- = 0$. By manipulating Eq.~\eqref{eq:QuadConst} properly, we arrive at
\begin{align}
\hat{E}^2 &= \frac{\kappa \rho + 2\epsilon\tilde\sigma \pm 2 \sqrt{\tilde\sigma(\tilde\sigma\epsilon^2 + \rho \epsilon \kappa - \eta \kappa^2)}}{\rho^2+4\eta\tilde\sigma} \label{eq:energy1} \; ,\\
\hat{J}_z &= \frac{\rho \hat{E}^2 - \kappa}{2\sigma \hat{E}} \label{eq:angmom1} \; ,
\end{align}
where
\begin{align}
\kappa &= d_1 h_2 - d_2 h_1 \; , \nonumber\\
\epsilon &= d_1 g_2 - d_2 g_1 \; , \nonumber\\
\rho &= f_1 h_2 - f_2 h_1 \; , \label{eq:determinants}\\
\eta &= f_1 g_2 - f_2 g_1 \; , \nonumber\\
\tilde\sigma &= g_1 h_2 - g_2 h_1 \;  \nonumber
\end{align}
are the determinants appearing in \cite{Schmidt:2002}. Thanks to the identity $\epsilon\rho - \kappa\eta = \tilde{\sigma} \zeta$, where 
\begin{equation}
\zeta = d_1 f_2 - d_2 f_1 \; ,
\end{equation}
we can rearrange Eq. \eqref{eq:energy1} as
\begin{equation} \label{eq:energy2}
    \hat{E}^2 = \frac{\kappa\rho + 2\epsilon\tilde{\sigma} - 2 \sgn{(\hat{J}_z)}\, \tilde{\sigma}\sqrt{\epsilon^2 + \kappa \zeta}}{\rho^2 + 4\eta\tilde{\sigma}} \, .
\end{equation}
Since for $\hat{a} = \sigma = 0$ the determinant $\tilde{\sigma} = 0$ and the Eq.~\eqref{eq:angmom1} is singular, it is better to substitute $\hat{E}^2$ into Eq. \eqref{eq:angmom1} and rearrange it as follows
\begin{equation} \label{eq:angmom2}
    \hat{J}_z = \frac{\epsilon\rho - 2\kappa\eta - \sgn{(\hat{J}_z)}\,\rho \sqrt{\epsilon^2 + \kappa \zeta}}{(\rho^2 + 4\eta\tilde{\sigma}) \hat{E}}\; .
\end{equation}
The signs of $\hat{J}_z$ appearing in Eqs.~\eqref{eq:energy2} and \eqref{eq:angmom2} have been numerically verified for spin values $\abs{\sigma} \leq 1$.

The constants of motion $\hat{E}$ and $J_z$ for given $p$ and $e$ have two solutions corresponding to the corotating orbit and the counterrotating orbit. We can choose the coordinates such that the $z$ axis is parallel to the total angular momentum, i.e. $\hat{J}_z>0$. This convention implies that $\ha>0$ corresponds to corotating orbits and $\ha<0$ to counterrotating orbits. The spins of the secondary particle and of the central black hole are parallel when $\ha\,\sigma>0$ and antiparallel when $\ha\,\sigma<0$.

For $e=0$, both the numerator and the denominator of Eq.~\eqref{eq:energy2} become zero. This inconvenience can be avoided by noticing that a coefficient $e$ can be factored out from the determinants \eqref{eq:determinants} and canceled out in Eq.~\eqref{eq:energy2}. In this fashion, the solution~\eqref{eq:energy2} is valid even for $e=0$. Actually, this allows us to verify that for $e=0$ Eqs.~\eqref{eq:energy2} and \eqref{eq:angmom2} are identical to Eqs.~(59) and (60) given in \citep{Piovano:2020}.

\begin{figure}[!ht]
  \centering  
  \includegraphics[width=0.48\textwidth]{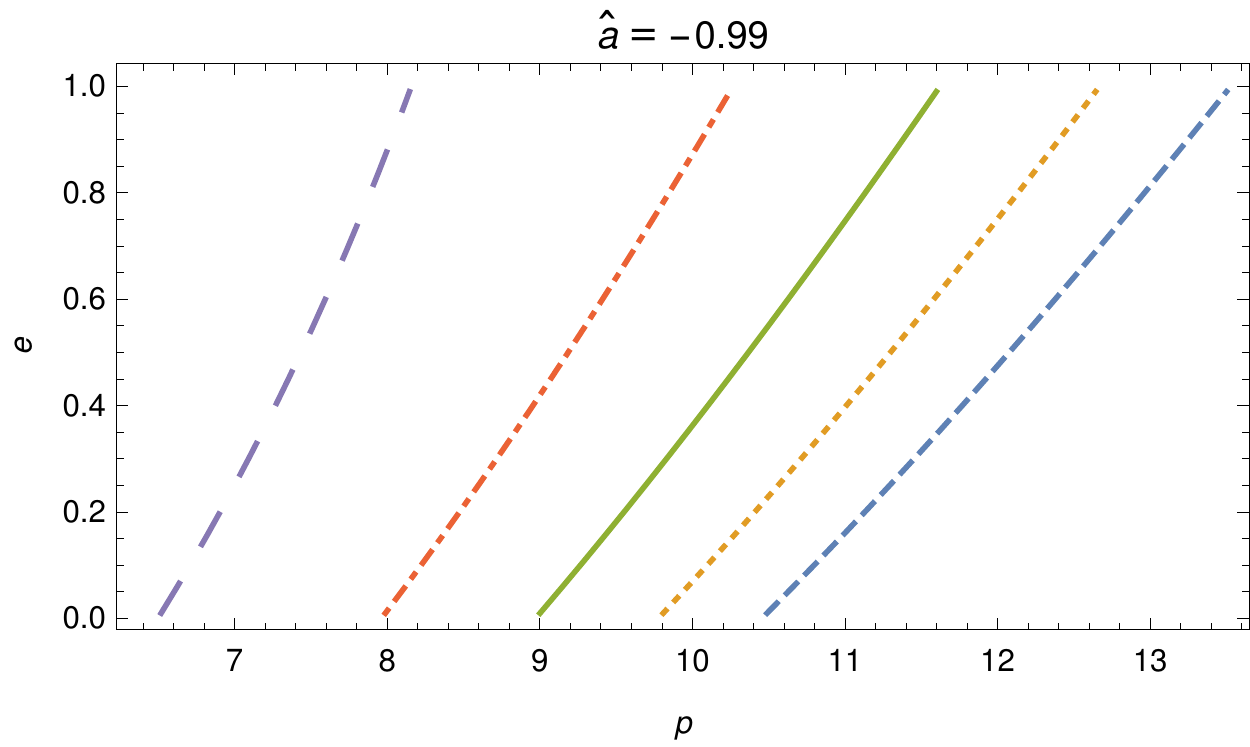} 
  \includegraphics[width=0.48\textwidth]{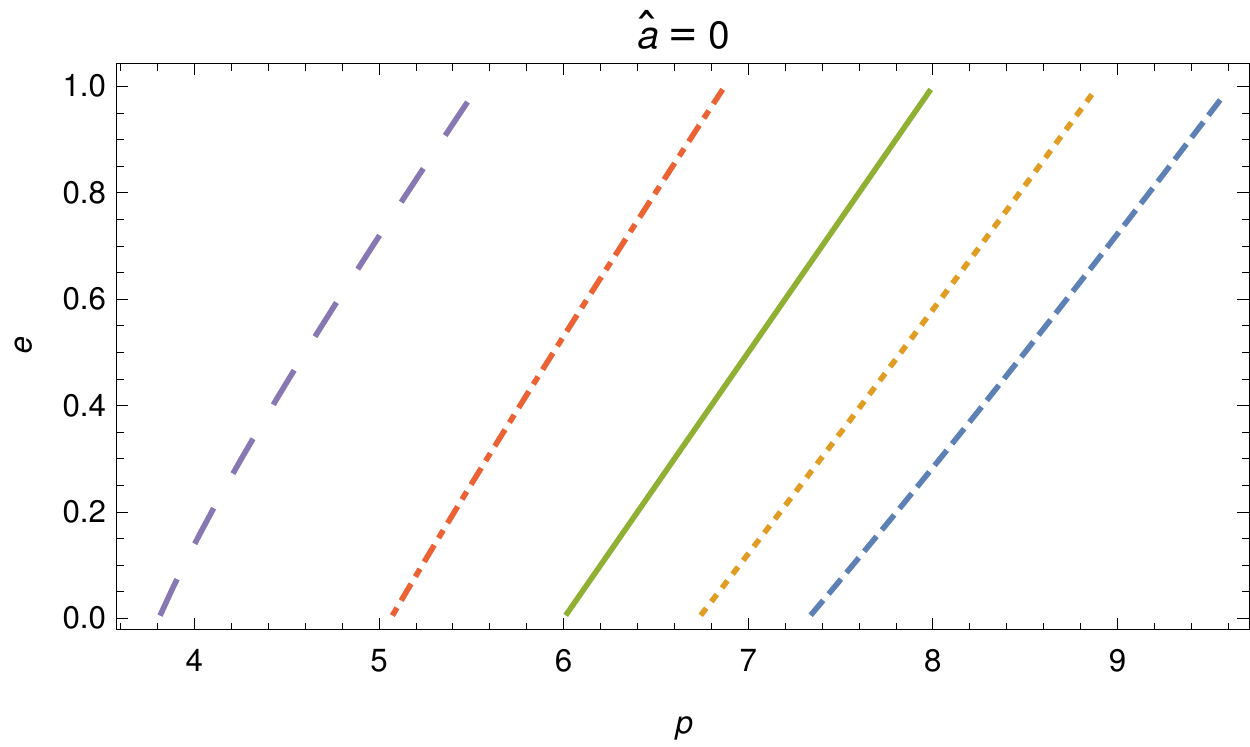} 
  \includegraphics[width=0.48\textwidth]{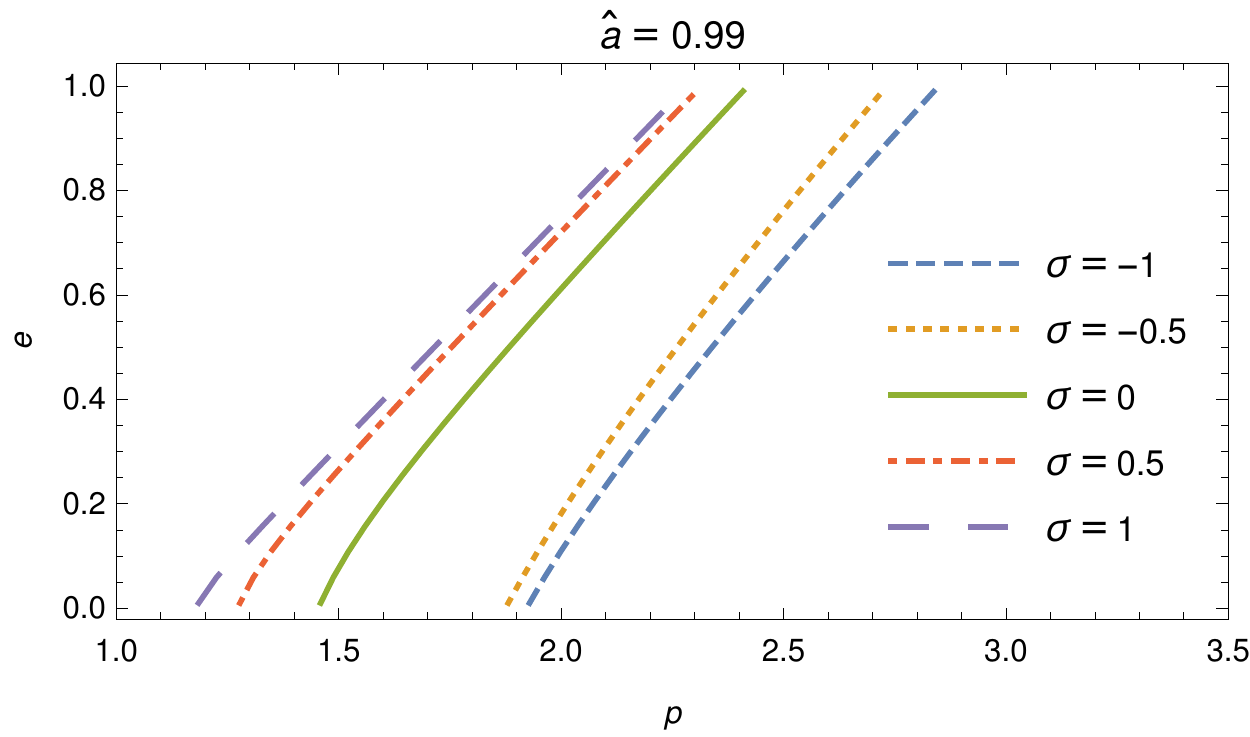} 
  \caption{Separatrices for different Kerr parameters and spins. Points ($p,~e$) on the depicted lines correspond to orbits asymptotically approaching the unstable circular orbit lying at $\hr = p/(1+e)$. For given $e$ the semi-latus rectum $p$ of the separatrix decreases with increasing spin. Therefore, for a spinning particle it is possible to approach the horizon closer than a nonspinning particle. Note that even though the EMRI relevant values of the secondary spin are $\sigma \leq q \ll 1$, we use much higher spin values to make the differences between the separatrices more prominent and visible. All plots are for $\hat{J}_z>0$. }
  \label{fig:separatrix}
\end{figure}

There is a limit between the bounded and unbounded equatorial orbits defined by a separatrix. The term unbounded orbits includes orbits escaping to infinity and orbits plunging to the central black hole. In the case the separatrix splits plunging and bounded orbits, it holds that $R'_\sigma(\hr_1) = 0$ and $R'_\sigma(\hr_2) < 0$. The orbit with $R'_\sigma(\hr_1) = 0$ is an unstable circular orbit, while a trajectory originating from $\hr_2$ with energy and angular momentum satisfying Eqs.~\eqref{eq:energy2} and \eqref{eq:angmom2} will asymptotically approach the circular orbit at $\hr_1$ either evolved forward or backward in time.\footnote{In the limiting case that $\hr_1=\hr_2$ the orbit is circular $(e=0)$ and marginally stable, since it holds that $R_\sigma(\hr_1) = R'_\sigma(\hr_1) = R''_\sigma(\hr_1) = 0 $. This orbit is often called the innermost stable circular orbit (ISCO).} For a given Kerr parameter $\ha$ and spin $\sigma$ the effective potential $R_\sigma$ depends on $\hat{E}(p,e)$ and $\hat{J}_z(p,e)$, therefore the separatrices can be plotted on the $p-e$ plane splitting it into two parts. In one part of the plane lie the bounded orbits, while in the other part lie unbounded orbits or initial conditions, which do not correspond to an orbit (Fig. \ref{fig:separatrix}). We can see that for given $e$ the semi-latus rectum $p$ of the separatrix decreases with increasing spin. 

\begin{figure}[!ht]
  \centering  
  \includegraphics[width=0.48\textwidth]{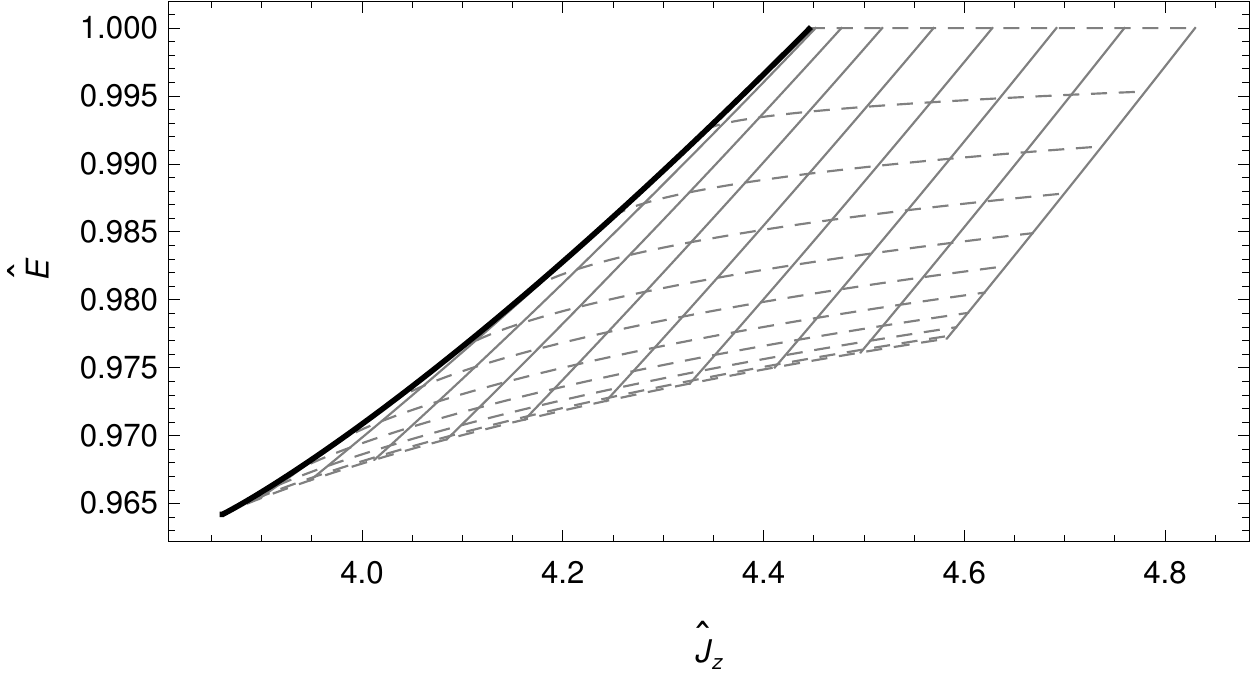} 
  \includegraphics[width=0.48\textwidth]{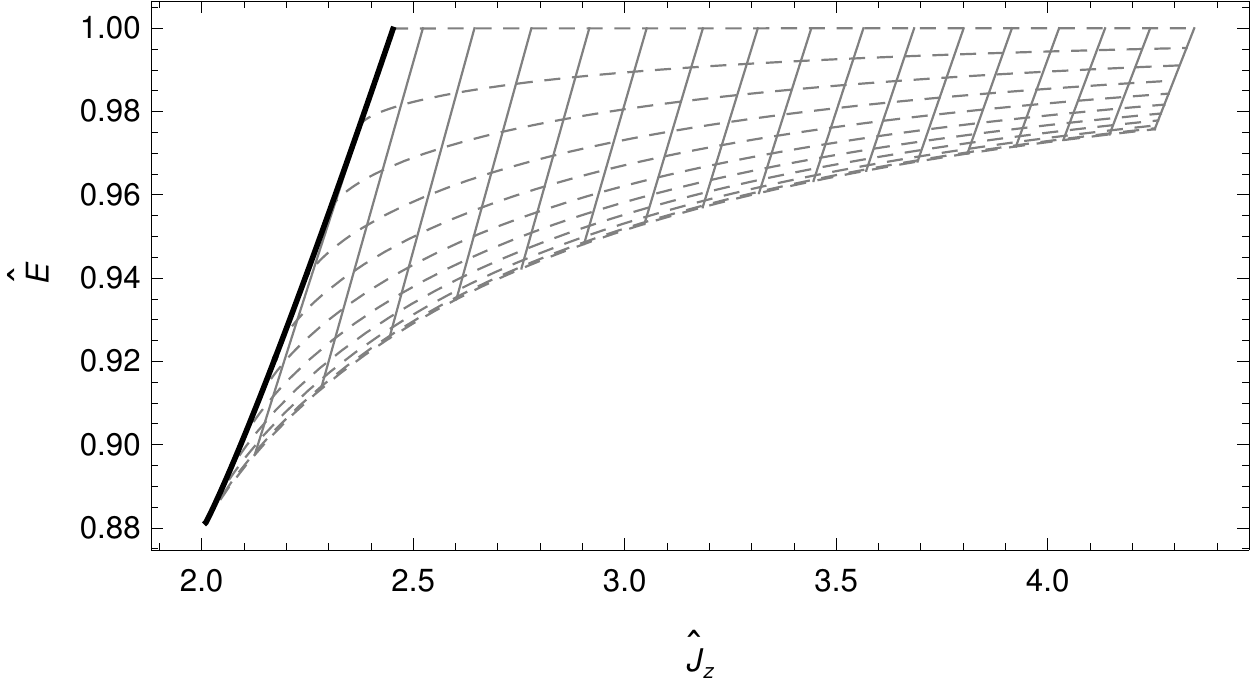} 
  \caption{Separatrices (black thick solid) in the $\hat{J}_z-\hat{E}$ plane along with lines of constant semi-latus rectum (grey solid) and eccentricity (grey dashed) for Kerr parameter $\ha=-0.5$ (top panel) and $\ha=0.5$ (bottom panel). In both case the secondary spin is $\sigma=0.5$. The eccentricity lines start at $e=0$ for lower energies and reach $e=1$ when $\hat{E} = 1$ with step $0.1$.
  The semi-latus rectum ranges from $p=10$ to $p=20$ for $\ha=-0.5$ and from $p=3$ to $p=20$  for $\ha=0.5$ with step $1$ in both plots. At a separatrix the semi-latus rectum is the lowest and is increasing with increasing $\hat{J}_z$.}
  \label{fig:separatrix_je}
\end{figure}

Figure~\ref{fig:separatrix_je} shows two cases of a separatrix on the $\hat{J}_z-\hat{E}$ plane along with a grid of constant $p$ and $e$ lines. Note that the intersection point between the separatrix and the line $e=0$ lying at the left lower corner of both panels of Fig.~\ref{fig:separatrix_je} represents ISCO. 

\subsection{Frequencies of eccentric equatorial orbits}

The radial motion of a particle in the equatorial plane parametrized by the time parameter $\lambda$ has a period $\Lambda_r$. This period can be defined as the time needed to go from the apocenter to the pericenter and back. Hence, $\Lambda_r$ can be found by integrating the inversion of Eq.~\eqref{eq:EOM_r}, i.e.,
\begin{equation}
    \frac{\rmd \lambda}{\rmd \hr} = \frac{1}{\sqrt{R_\sigma(\hr)}} \, , \label{eq:EOM_r2}\\
\end{equation}
over the above two branches (first from $\hr_1$ to $\hr_2$ and then from $\hr_2$ to $\hr_1$) with respect to the radius $\hr$. However, the integration over one branch is equal to the integration over the other. Hence, we can find the $\Lambda_r$ by integrating Eq.~\eqref{eq:EOM_r} over the first branch to obtain the time elapsed during the first branch and multiply the result by two \citep{Fujita:2009}, i.e.,
\begin{equation} \label{eq:Lambda_r}
    \Lambda_r = 2\int_{\hr_1}^{\hr_2} \frac{\rmd \hr}{\sqrt{R_\sigma(\hr)}} \; .
\end{equation}
The radial frequency can be defined as $\Upsilon_r = 2\pi/\Lambda_r$. If we set the initial radius to $r(\lambda=0) = r_1$, then the radius $r(\lambda)$ is an even function and can be written as
\begin{equation} \label{eq:r_lambda}
    r(\lambda) = r^{(0)} + \sum_{n=1}^\infty r^{(n)} \cos(n \Upsilon_r \lambda) \; .
\end{equation}

After substituting Eq.~\eqref{eq:r_lambda} to Eqs. \eqref{eq:EOM_t} and \eqref{eq:EOM_phi} and  integrating them, we obtain
\begin{align}
    \hat{t}(\lambda) &= \Gamma \lambda + \Delta \hat{t}(\lambda) \; , \nonumber \\
    \phi(\lambda) &= \Upsilon_\phi \lambda + \Delta \phi(\lambda) \; , \label{eq:t_phi_lambda}
\end{align}
where $\Gamma$ and $\Upsilon$ are frequencies with respect to $\lambda$ and functions $\Delta \hat{t}(\lambda)$ and $\Delta \phi(\lambda)$ are periodic with period $\Lambda_r$. Note that since the function $\hr(\lambda)$ is even, the functions $V^t(\hr(\lambda))$ and $V^\phi(\hr(\lambda))$ are even in $\lambda$ as well. Hence, after the aforementioned integration and the subtraction of the linear term $\Gamma\lambda$ or $\Upsilon_\phi\lambda$, respectively in Eq.~\eqref{eq:t_phi_lambda}, the functions $\Delta \hat{t}(\lambda)$ and $\Delta \phi(\lambda)$ are odd and can be written as series of sines.

The average rate of change of the azimuthal coordinate and time with respect to $\lambda$ is
\begin{align}
    \Upsilon_\phi &= \frac{2}{\Lambda_r} \int_{\hr_1}^{\hr_2} \frac{V^\phi(\hr)}{\sqrt{R_\sigma(\hr)}} \rmd \hr \; , \label{eq:Upsilon_phi}\\
    \Gamma &= \frac{2}{\Lambda_r} \int_{\hr_1}^{\hr_2} \frac{V^t(\hr)}{\sqrt{R_\sigma(\hr)}} \rmd \hr \; . \label{eq:Gamma}
\end{align}
These integrals can be solved in terms of Lauricella's hypergeometric functions \citep{Hackmann:2014tga}. However, for achieving this, the exact values of the roots of the radial potential $\hr^4 R_\sigma(\hr)$, which is eighth order polynomial in $\hr$, must be found. This task can be only performed numerically. Thus, instead the integrals \eqref{eq:Lambda_r}, \eqref{eq:Upsilon_phi} and \eqref{eq:Gamma} were calculated directly numerically. These integrals have singular points at $\hr_1$ and $\hr_2$,  but this difficulty can be overcome. Namely, first we factor out the roots
\begin{equation}
    R_\sigma(\hr) = (\hr-\hr_1)(\hr_2-\hr)Q(\hr) \; ,
\end{equation}
where $\hr^4 Q(\hr)$ is sixth order polynomial. To remove the singularities, an angle like coordinate $\chi \in \left[0,\pi\right)$ is used by applying the transformation
\begin{equation}
    \hr = \frac{p}{1+e\cos\chi} \; .
\end{equation}
Then, the integrals take the form
\begin{align}
    \Lambda_r &= \frac{2\sqrt{1-e^2}}{p} \int_{0}^{\pi} \frac{1}{\sqrt{J(\chi)}} \rmd \chi \; , \\
    \Upsilon_\phi &= \frac{2\sqrt{1-e^2}}{\Lambda_r p} \int_{0}^{\pi}  V^\phi\left(\frac{p}{1+e\cos\chi}\right) \frac{1}{\sqrt{J(\chi)}} \rmd \chi \; , \\
    \Gamma &= \frac{2\sqrt{1-e^2}}{\Lambda_r p} \int_{0}^{\pi}  V^t\left(\frac{p}{1+e\cos\chi}\right) \frac{1}{\sqrt{J(\chi)}}  \rmd \chi \; ,
\end{align}
where 
\begin{equation}
 J(\chi) = \sum_{k=0}^6 (1+e\cos\chi)^k \sum_{l=0}^k \frac{j^{(p)}_l j^{(e)}_{k-l}}{(1-e^2)^{k-l} p^l}
\end{equation}
is a polynomial in $\cos\chi$ with coefficients
\begin{align*}
 j^{(p)}_0 &= 1-\hat{E}^2 \; , \\
 j^{(p)}_1 &= -2 \; , \\
 j^{(p)}_2 &= \ha^2+2 \ha \hat{E} x+x^2 \; , \\
 j^{(p)}_3 &= -2((1-\hat{E}^2) \sigma^2 - \hat{E} \sigma x + x^2) \; , \\
 j^{(p)}_4 &= 4\sigma^2 \; , \\
 j^{(p)}_5 &= -2 \ha \sigma (\ha \sigma+x (\hat{E}\sigma +x)) \; , \\
 j^{(p)}_6 &= \sigma^2 ((1 - \hat{E}) \sigma - x) ((1 + \hat{E}) \sigma + x)
\end{align*}
and
\begin{align*}
 j^{(e)}_0 &= 1 \; , \\
 j^{(e)}_1 &= 2 \; , \\
 j^{(e)}_2 &= e^2+3 \; , \\
 j^{(e)}_3 &= 4 (e^2+1) \; , \\
 j^{(e)}_4 &= e^4+10 e^2+5 \; , \\
 j^{(e)}_5 &= 2 (e^2+3) (3 e^2+1) \; , \\
 j^{(e)}_6 &= e^6+21 e^4+35 e^2+7 \; .
\end{align*}
The polynomial $J(\chi)$ for $\sigma=0$ is identical to the polynomial~(40) in \citep{Schmidt:2002} with Carter constant $Q=0$ up to the factor $1-e^2$ due to a different definition of $J(\chi)$ used in \citep{Schmidt:2002}.

We can define the frequencies with respect to the coordinate time as
\begin{align}
    \hat{\Omega}_r &= \frac{\Upsilon_r}{\Gamma} = \frac{\pi p}{\sqrt{1-e^2} \int_0^\pi V^t(\hr(\chi))/\sqrt{J(\chi)} \rmd \chi} \; , \label{eq:Omega_r}\\
    \hat{\Omega}_\phi &= \frac{\Upsilon_\phi}{\Gamma} = \frac{\int_0^\pi V^\phi(\hr(\chi))/\sqrt{J(\chi)} \rmd \chi}{\int_0^\pi V^t(\hr(\chi))/\sqrt{J(\chi)} \rmd \chi} \; . \label{eq:Omega_phi}
\end{align}
We have numerically verified the above frequency formulas by comparing them with frequencies obtained by a direct integration of the MPD equations for the respective eccentric orbits. To integrate the MPD equations an implicit Gauss-Runge-Kutta integrator was used as described in \citep{Lukes-Gerakopoulos:2014dma}.

The equatorial plane equations of motion~\eqref{eq:EOM_equatorial} given in $t$, $r$ and $\phi$ can be rewritten in $\lambda$, $t$ and $\phi$ parametrized by $\chi$, i.e.
\begin{align}
 \frac{\rmd\lambda}{\rmd\chi} &= \sqrt{\frac{1-e^2}{p^2 J(\chi)}} \label{eq:EOM_lambda} \\
 \frac{\rmd \hat{t}}{\rmd\chi} &= V^t\left(\frac{p}{1+e\cos\chi}\right) \sqrt{\frac{1-e^2}{p^2 J(\chi)}} \label{eq:EOM_t2} \\
 \frac{\rmd\phi}{\rmd\chi} &= V^\phi\left(\frac{p}{1+e\cos\chi}\right) \sqrt{\frac{1-e^2}{p^2 J(\chi)}} \label{eq:EOM_phi2}
\end{align}
These equations will be used later on, when the energy and angular momentum fluxes are calculated.

\section{Gravitational wave fluxes}
\label{sec:GWfluxes}

\subsection{Teukolsky formalism}

To calculate the GW fluxes we employ the Teukolsky formalism. The GWs are described perturbatively using the Weyl curvature scalar
\begin{equation}
    \Psi_4 = -C_{\alpha\beta\gamma\delta} n^\alpha \mbar^\beta n^\gamma \mbar^\delta \, ,
\end{equation}
where $n^\mu$ and $\mbar^\mu$ are components of the Kinnersley tetrad
\begin{align}
    n^\mu &= \frac{1}{2\Sigma}\left( \varpi^2 , - \Delta , 0, a \right) \; , \\
    \mbar^\mu  &= \frac{\rho}{\sqrt{2}} \left( ia\sin\theta , 0,  -1,   i\csc\theta \right)\; ,
\end{align}
where $\rho = -(r-ia\cos\theta)^{-1}$. The Weyl scalar $\Psi_4$ is zero for the Kerr spacetime and its perturbation is governed by the TE
\begin{equation} \label{eq:teuk}
    {}_s\mathcal{O} \, {}_s\psi(t,r,\theta,\phi) = 4\pi \Sigma T
\end{equation}
with spin weight $s=-2$ for $_{-2}\psi = \rho^{-4} \Psi_4$ in the case of the GWs \citep{Teukolsky:1973ha}. 

\subsubsection{Frequency domain approach}
\label{sec:fd}

This partial differential equation can be separated into ordinary differential equations after a Fourier transform in $t$ and $\phi$
\begin{equation} \label{eq:psi_fourier}
    {}_{-2}\psi = \sum_{l,m}^{\infty} \frac{1}{2\pi} \int_{-\infty}^{\infty} \rmd \omega e^{-i\omega t} \psi_{lm\omega}(r) {}_{-2}S_{lm}^{a\omega}(\theta,\phi) \; ,
\end{equation}
where $_{-2}S_{lm}^{a\omega}(\theta, \phi)$ is spin weighted spheroidal harmonic function with spin weight $-2$ normalized as
\begin{equation}
    \int \rmd \Omega \left| _{-2}S_{lm}^{a\omega}(\theta, \phi) \right| = 1 \; .
\end{equation}
For simplicity we use the notation $S_{lm}^{a\omega}(\theta) = {}_{-2}S_{lm}^{a\omega}(\theta,0)$ for the angular part henceforth. To calculate the angular function the \textit{Black Hole Perturbation Toolkit} \citep{BHPToolkit} has been employed.

After the separation, an ordinary differential equation 
\begin{equation} \label{eq:teuk_radial}
    \mathcal{D}\psi_{lm\omega}(r) = \mathcal{T}_{lm\omega} 
\end{equation}
is obtained for the radial part $\psi_{lm\omega}(r)$, where $\mathcal{D}$ is a differential operator that can be found, e.g., in \citep{Teukolsky:1973ha} and $\mathcal{T}_{lm\omega}$ is a source term discussed below. The asymptotic behavior of the homogeneous solutions $R_{lm\omega}(r)$ of Eq. \eqref{eq:teuk_radial} is discussed in \citep{Drasco:2005kz, Mino:1997bx}. To satisfy physical boundary conditions, the solution must be purely outgoing at infinity and purely ingoing at the horizon; in other words, we are dealing with a retarded solution. We will denote a homogeneous solution satisfying the first condition as $R^+_{lm\omega}$ and a solution satisfying the second condition as $R^-_{lm\omega}$.\footnote{These functions are often denoted $R^\infty_{lm\omega}$ and $R^{\rm H}_{lm\omega}$ or $R^{\rm Up}_{lm\omega}$ and $R^{\rm In}_{lm\omega}$.} An inhomogeneous solution satisfying boundary conditions can be found using the Green function formalism as
\begin{equation} \label{eq:psi_radial}
    \psi_{lm\omega}(r) = C^+_{lm\omega}(r) R^+_{lm\omega}(r) + C^-_{lm\omega}(r) R^-_{lm\omega}(r) \; ,
\end{equation}
where the amplitudes are
\begin{equation} \label{eq:Cpm_lmomega}
    C^\pm_{lm\omega}(r) = \frac{1}{W} \int_{r_+}^\infty \Theta^\pm(r,r') \frac{R^\mp_{lm\omega}(r') \mathcal{T}_{lm\omega}(r')}{\Delta^2(r')} \rmd r'
\end{equation}
with the invariant Wronskian
\begin{equation}
    W = \frac{R^+_{lm\omega}(r) \partial_r R^-_{lm\omega}(r) - (\partial_r R^+_{lm\omega}(r)) R^-_{lm\omega}(r)}{\Delta(r)}
\end{equation}
and the Heaviside step functions defined as
\begin{equation}
    \Theta^+(r,r') = \Theta(r'-r) \; , \qquad \Theta^-(r,r') = \Theta(r-r') \; .
\end{equation}
Since we are interested in GW fluxes at the horizon and at infinity, we will denote the relevant amplitudes as $C^-_{lm\omega} \equiv C^-_{lm\omega}(r\rightarrow r_+)$ and $C^+_{lm\omega} \equiv C^+_{lm\omega}(r\rightarrow \infty)$ respectively. In fact, the amplitudes are constant for $r<r_1$ and $r>r_2$.

The source term in \eqref{eq:teuk_radial} can be written as
\begin{equation} \label{eq:source_Tlmomega}
    \mathcal{T}_{lm\omega} = \int \rmd t \, \rmd \theta \, \rmd \phi  \, \Delta^2 (\mathcal{T}_{nn} + \mathcal{T}_{n\mbar} + \mathcal{T}_{\mbar\mbar}) e^{i\omega t - im\phi} \, ,
\end{equation}
where 
\begin{align}
    \mathcal{T}_{nn} &= f_{nn}^{(0)}(r,\theta) \sqrt{-g} T_{nn} \; , \nonumber\\
    \mathcal{T}_{n\mbar} &= \partial_r(f_{n\mbar}^{(1)}(r,\theta) \sqrt{-g} T_{n\mbar}) \nonumber\\ & \;\;\;\;\; + f_{n\mbar}^{(0)}(r,\theta) \sqrt{-g} T_{n\mbar} \; , \label{eq:source_Tnm}\\
    \mathcal{T}_{\mbar\mbar} &= \partial_{rr}(f_{\mbar\mbar}^{(2)}(r,\theta) \sqrt{-g} T_{\mbar\mbar}) + \nonumber\\ & \;\;\;\;\; \partial_r(f_{\mbar\mbar}^{(1)}(r,\theta) \sqrt{-g} T_{\mbar\mbar}) + f_{\mbar\mbar}^{(0)}(r,\theta) \sqrt{-g} T_{\mbar\mbar} \; . \nonumber
\end{align}
The functions $f_{ab}^{(i)}(r,\theta)$ can be found in \citep{Piovano:2020}. Projections of the stress energy tensor onto a tetrad $e^{(a)}_\mu$ read
\begin{multline} \label{eq:T_ab}
    T_{ab} = \frac{1}{\sqrt{-g}} \left( C_{ab}^0 - C_{ab}^\sigma \right) \delta^3 \\ - \frac{1}{\sqrt{-g}} \partial_\rho \left((v^t)^{-1} S^{\rho (\mu} v^{\nu )} \delta^3\right) e_{\mu}^{(a)} e_{\nu}^{(b)}  \; ,
\end{multline}
where
\begin{align}
    C_{ab}^{0} &= (v^t)^{-1} P^{(\mu} v^{\nu)} e_{\mu}^{(a)} e_{\nu}^{(b)} \; , \nonumber\\
    C_{ab}^{\sigma} &= (v^t)^{-1} S^{\rho(\mu} \Gamma^{\nu)}{}_{\rho\lambda} v^\lambda e_{\mu}^{(a)} e_{\nu}^{(b)} \, .\label{eq:Cab}
\end{align}
 The four-vectors $P^\mu$ and $v^\mu$ as well as the spin tensor $S^{\mu\nu}$ are functions of time, the Christoffel symbols are evaluated at the coordinates of the particle $r_p(t),\theta_p(t)$, the delta functions are functions of both the space coordinates $r, \theta, \phi$ and the coordinate time $t$ and the square root of the determinant $\sqrt{-g}$, the functions $f_{ab}^{(i)}$ and the tetrad legs $e^{(a)}_\mu$ are functions of $r$ and $\theta$. In our case, $e^{(a)}_\mu$, $e^{(b)}_\mu$ are the Kinnersley tetrad components $n_\mu$ and $\mbar_\mu$. 

After integrating Eq.~\eqref{eq:source_Tlmomega} over $\theta$ and $\phi$ and Eq.~\eqref{eq:Cpm_lmomega} over $r$ using rules for integrating delta function, we obtain a relation for the amplitudes
\begin{equation}
    C^\pm_{lm\omega} = \int_{-\infty}^{\infty} \rmd t e^{i\omega t - i m \phi_p(t)} I^\pm_{lm\omega}(r_p(t),\theta_p(t))
\end{equation}
where
\begin{multline} \label{eq:Ipm}
    I^\pm_{lm\omega}(r,\theta) = \frac{1}{W} \left(A_0 - (A_1+B_1) \frac{\rmd}{\rmd r} \right. \\ \left. + (A_2+B_2) \frac{\rmd^2}{\rmd r^2} - B_3 \frac{\rmd^3}{\rmd r^3} \right) R^{\mp}_{lm\omega}(r) \; .
\end{multline}
The coefficients $A_i$ in their general form can be found in Appendix~\ref{sec:coefficientsAi}. 

Up to this point the derivation of GW fluxes holds for a generic orbit of a spinning particle. In the following part, we confine it to equatorial orbits with the spin parallel to the $z$ axis as described in Sec. \ref{sec:equatorial_dynamics}. 

Thanks to the fact that the quantity $I^\pm_{lm\omega}(r_p(t),\pi/2)e^{im(\Omega_\phi t - \phi_p(t))}$ is periodic in time with frequency $\Omega_r$ (see eg. \citep{Glampedakis:2002ya} for details), we can write the amplitude as a sum over discrete frequencies
\begin{align}
    C^\pm_{lm\omega} &= \sum_{n=-\infty}^\infty C^\pm_{lmn} \delta(\omega - \omega_{mn}) \; , \label{eq:C_lmomega}\\ 
    \omega_{mn} &= m\Omega_\phi + n\Omega_r \; .
\end{align}
The partial amplitudes can be calculated as Fourier coefficients by integrating over one period $T_r=2\pi/\Omega_r$
\begin{multline}
    C^\pm_{lmn} = \Omega_r \int_0^{T_r} \rmd t I^\pm_{lm\omega_{mn}}(r_p(t),\pi/2) \\ \times \exp(i\omega_{mn} t - i m \phi_p(t))\; .
\end{multline}
However, it is more convenient to integrate over the time parameter $\lambda$
\begin{multline}
    C^\pm_{lmn} = \Omega_r \int_0^{\Lambda_r} \rmd \lambda \frac{\rmd t}{\rmd \lambda} I^\pm_{lm\omega_{mn}}(r_p(t(\lambda)),\pi/2) \\ \times \exp(i\omega_{mn} t(\lambda) - i m \phi_p(\lambda))\; . \label{eq:integral_lambda_Cpm}
\end{multline}

The integration over the two branches of the motion (from $r_1$ to $r_2$ which correspond to $\lambda$ from $0$ to $\Lambda_r/2$ and from $r_2$ to $r_1$ which correspond to $\lambda$ from $\Lambda_r/2$ to $\Lambda_r$) differs only by the sign of the radial velocity. Therefore, we can break the integral to two integrals, the first from $0$ to $\Lambda_r/2$ and the second from $\Lambda_r$ to $\Lambda_r/2$ (note the reverse direction of integration). Using the identities \eqref{eq:t_phi_lambda} we can write
\begin{equation}
    \omega_{mn}t(\lambda) - m \phi(\lambda) = n \Upsilon_r \lambda + \omega_{mn} \Delta t - m \Delta \phi \, .
\end{equation}
From the fact that $\Delta t$ and $\Delta\phi$ are series of sines with period $\Lambda_r$, it holds $\Delta t(\Lambda_r-\lambda) = -\Delta t(\lambda)$ and $\Delta \phi(\Lambda_r-\lambda) = -\Delta \phi(\lambda)$. After changing  the integration variable to $\chi$, we can write the integral as a sum over the sign $D_r=\pm$ of the radial velocity, on which the coefficients $A_{i}$ depend, i.e. 
\begin{multline} \label{eq:Cpm_lmn_final}
    C^\pm_{lmn} = \Omega_r \int_0^\pi  \rmd \chi \frac{\rmd \lambda}{\rmd \chi} \sum_{D_r = \pm} \frac{\rmd t}{\rmd \lambda} I^\pm_{lm\omega_{mn}}(r(\chi),\pi/2,D_r) \\ \times \exp(i D_r ( \omega_{mn} t(\chi) - m \phi(\chi) ) ) \; ,
\end{multline}
where $\rmd \lambda / \rmd \chi$ comes from Eq.~\eqref{eq:EOM_lambda}, $I^\pm_{lm\omega_{mn}}$ comes from Eq.~\eqref{eq:Ipm} and $t(\chi)$, $\phi(\chi)$ are calculated from Eqs.~\eqref{eq:EOM_t2},~\eqref{eq:EOM_phi2}.

The metric perturbation $h_{\mu\nu} = \mathcal{O}(q)$ which can be defined as $g^{\rm exact}_{\mu\nu} = g_{\mu\nu} + h_{\mu\nu} + \mathcal{O}(q^2)$, can be calculated from the Weyl scalar $\Psi_4$ \citep{Chrzanowski:1975}. GWs consist of two polarizations and the metric perturbation can be written as $h_{\mu\nu} = h_+ e_{\mu\nu}^+ + h_\times e_{\mu\nu}^\times$ where $e_{\mu\nu}^+$ and $e_{\mu\nu}^\times$ are the polarization tensors. At infinity, the relation between the strain $h = h_+ - i h_\times$ and the Weyl scalar is 
\begin{align}
  \Psi_4(r\rightarrow\infty) = \ddot{h}/2 \, ,  
\end{align}
where the dots denote derivative with respect to the BL coordinate time $t$. From Eqs.~\eqref{eq:psi_fourier},~\eqref{eq:psi_radial},~\eqref{eq:C_lmomega} and the asymptotic behavior of $R^+_{lm\omega}$ it holds
\begin{equation} \label{eq:strain}
    h = -\frac{2}{r} \sum_{lmn} \frac{C^+_{lmn}}{\omega_{mn}^2} S_{lm}^{a\omega_{mn}}(\theta) e^{-i\omega_{mn}(t-r^\ast) + im\phi} \; ,
\end{equation}
where $r^\ast$ is tortoise coordinate defined as $\rmd r^\ast/\rmd r = \varpi^2/\Delta$. The stress-energy tensor of the GW can be reconstructed from the strain which yields the energy and angular momentum fluxes at infinity
\begin{align}
    \left\langle \frac{\rmd E^\infty}{\rmd t} \right\rangle &= \sum_{l=2}^\infty \sum_{m=-l}^l \sum_{n=-\infty}^\infty \frac{\left|C^+_{lmn}\right|^2}{4\pi \omega_{mn}^2} \; , \\
    \left\langle \frac{\rmd J_z^\infty}{\rmd t} \right\rangle &= \sum_{l=2}^\infty \sum_{m=-l}^l \sum_{n=-\infty}^\infty \frac{m\left|C^+_{lmn}\right|^2}{4\pi \omega_{mn}^3} 
\end{align}
where the brackets denote time averaging. In the equatorial case, the average can be calculated over one period $T_r$. Similar derivation can be made for the fluxes at the horizon \citep{Teukolsky:1974yv}
\begin{align}
    \left\langle \frac{\rmd E^{\rm H}}{\rmd t} \right\rangle &= \sum_{l=2}^\infty \sum_{m=-l}^l \sum_{n=-\infty}^\infty \alpha_{lmn}\frac{\left|C^-_{lmn}\right|^2}{4\pi \omega_{mn}^2} \; ,\\
    \left\langle \frac{\rmd J_z^{\rm H}}{\rmd t} \right\rangle &= \sum_{l=2}^\infty \sum_{m=-l}^l \sum_{n=-\infty}^\infty \alpha_{lmn}\frac{m\left|C^-_{lmn}\right|^2}{4\pi \omega_{mn}^3} \; ,
\end{align}
where
\begin{equation}
    \alpha_{lmn} = \frac{256(2Mr_+)^5 P (P^2+3\epsilon^2) (P^2+16\epsilon^2)\omega_{mn}^3}{\left| \mathscr{C}_{lm\omega_{mn}} \right|^2}
\end{equation}
with $\epsilon = \sqrt{M^2-a^2}/(4Mr_+)$, $P=\omega_{mn} - ma/(2Mr_+)$ and the Teukolsky-Starobinsky constant is
\begin{widetext}
\begin{multline}
    \left| \mathscr{C}_{lm\omega} \right|^2 = \left( \left( \lambda_{lm\omega} + 2 \right)^2 + 4a\omega (m - a\omega) \right)\left( \lambda_{lm\omega}^2 + 36a\omega (m - a\omega) \right) \\ - \left( 2\lambda_{lm\omega} + 3 \right) \left( 48a\omega (m - 2a\omega) \right) + 144 \omega^2 \left( M^2-a^2 \right)\; .
\end{multline}
\end{widetext}

The partial amplitudes $C^\pm_{lmn}$ are proportional to the secondary mass $\mu$ and therefore, if we use dimensionless quantities on the rhs, we obtain
\begin{align} \label{eq:fluxes}
    \left\langle \frac{\rmd E^\infty}{\rmd t} \right\rangle &= q^2 \sum_{l,m,n} \frac{\left|\hat{C}^+_{lmn}\right|^2}{4\pi \hat{\omega}_{mn}^2} \equiv q^2 \sum_{l,m,n} \mathcal{F}^{E\infty}_{lmn} \; ,\\
    \left\langle \frac{\rmd J_z^\infty}{\rmd t} \right\rangle &= M q^2 \sum_{l,m,n} \frac{m \left|\hat{C}^+_{lmn}\right|^2}{4\pi \hat{\omega}_{mn}^3} \equiv M q^2 \sum_{l,m,n} \mathcal{F}^{J_z \infty}_{lmn} \; ,
\end{align}
where we have defined the dimensionless fluxes $\mathcal{F}^{E\infty}_{lmn}$ and $\mathcal{F}^{J_z \infty}_{lmn}$ that do not depend on the mass ratio $q$. The horizon fluxes $\mathcal{F}^{E{\rm H}}_{lmn}$ and $\mathcal{F}^{J_z {\rm H}}_{lmn}$ can be defined in a similar fashion. We can write the dimensionless energy and angular momentum loss as
\begin{align}
    \left\langle \frac{\rmd \hat{E}^\infty}{\rmd \hat{t}} \right\rangle &= q \sum_{l,m,n} \mathcal{F}^{E\infty}_{lmn} \; ,\\
    \left\langle \frac{\rmd \hat{J}_z^\infty}{\rmd \hat{t}} \right\rangle &= q \sum_{l,m,n} \mathcal{F}^{J_z \infty}_{lmn} \; .
\end{align}

These fluxes can be used for calculating the evolution of the orbital parameters $p$ and $e$ during an adiabatic approximation of an inspiral.

\subsubsection{Time domain approach}
\label{sec:td}

To verify the frequency domain calculations, we numerically solved the TE \eqref{eq:teuk} in the time domain. For this, we have employed the time domain solver \texttt{Teukode} which is described in \citep{Harms:2013ib, Harms:2014dqa, Nagar:2014kha}. \texttt{Teukode} uses the method of lines, i.e. finite differences in space and Runge-Kutta for evolution in time. Instead of using Kinnersley tetrad and BL coordinates, it solves TE using Campanelli tetrad \citep{Campanelli:2000nc} and hyperboloidal horizon-penetrating (HH) coordinates\footnote{In this section $\rho$ denotes the radial HH-coordinate.} $(\tau,\rho,\theta,\varphi)$ (for their definition see Eq.~(10) in \citep{Harms:2014dqa}). These coordinates reach future null infinity $\mathscr{I}^+$ (``scri'') and horizon at finite radial coordinate $\rho_S$ so no extrapolation is needed to extract GW fluxes at infinity. Another advantage is that the coordinate light speed at the boundaries vanishes, therefore, no numerical boundary condition must be imposed. After the decomposition into azimuthal $m$-modes $\psi = \sum_m \psi_m e^{im\varphi}$ the equation in $(2+1)$-dimensional form reads
\begin{multline}
    \left( C_{\tau\tau} \partial_{\tau}^2 + C_{\tau\rho} \partial_\tau \partial_\rho + C_{\rho\rho} \partial_\rho^2 + C_{\theta\theta} \partial_\theta^2 \right. \\ \left. + C_\tau \partial_\tau + C_\rho \partial_\rho + C_\theta \partial_\theta + C_0 \right) \psi_m = S_s \; ,
\end{multline}
where the coefficients $C_{\tau\tau}, C_{\tau\rho}, \ldots$ are functions of $\rho$ and $\theta$ and $S_s$ is the source term for spinning particle discussed in \citep{Harms:2015ixa}.

The source term consists of derivatives of delta functions up to third order. For accurate results proper representation of delta functions must be used. Approximation as Gaussian function and piecewise polynomials as described in \citep{Sundararajan:2007jg} were implemented to the \texttt{Teukode}. According to \citep{Harms:2014dqa}, piecewise polynomial approximation is more accurate for circular equatorial orbits and faster to calculate than Gaussian approximation, whereas calculations with Gaussian approximation are more stable when the particle is moving in $\rho$ or $\theta$ direction. The third derivative of the delta function, which is needed for spinning particle, was implemented only as Gaussian approximation in the previous works. In our work we introduced to \texttt{Teukode} an approach suggested in \citep{Walden:1999}, which describes slightly different formulas for piecewise polynomial approximation to construct delta function and its derivatives. \texttt{Teukode} has been tested extensively on circular equatorial orbits of a spinning particle in \citep{Harms:2015ixa,Harms:2016d,Lukes-Gerakopoulos:2017,Nagar:2019,Piovano:2020}, but in this work it is tested for the first time on eccentric equatorial orbits of a spinning particle.

\subsection{Numerical results}

This Section discusses our numerical calculations of GW fluxes in the frequency domain (as described in Sec.~\ref{sec:fd}) and compare them with time domain results obtained from the \texttt{Teukode} (Sec.~\ref{sec:td}). 

First we present our approach to calculate quantities related to an orbit for given parameters $\ha$, $\sigma$, $p$ and $e$. These quantities include the energy and the angular momentum from Eqs.~\eqref{eq:energy2} and \eqref{eq:angmom2} respectively, the orbital frequencies $\hat{\Omega}_r$ and $\hat{\Omega}_\phi$ from Eqs. \eqref{eq:Omega_r} and \eqref{eq:Omega_phi} respectively and the functions $\hat{t}(\chi)$ and $\phi(\chi)$ from Eqs. \eqref{eq:EOM_t2} and \eqref{eq:EOM_phi2} respectively. The integrals~\eqref{eq:Omega_r} and \eqref{eq:Omega_phi} were calculated numerically using methods built-in to \textit{Mathematica}. We used extended precision to 48 places, because high precision of the parameters $a$ and $\omega = m\Omega_\phi + n\Omega_r$ is needed for the calculation of the radial function $R^\pm_{lm\omega}$.

\begin{figure}[!ht]
  \centering  
  \includegraphics[width=0.48\textwidth]{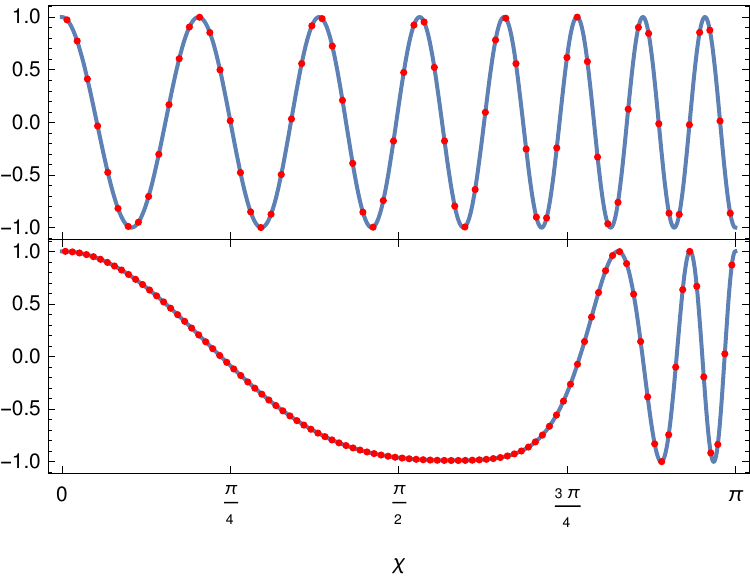}
  \caption{The real part of $\exp(i D_r(\omega_{mn}t(\chi) - m\phi(\chi)))$ for orbital parameters $\ha=0.9$, $\sigma=-0.5$, $p=12$, $e=0.2$ and $m=2$, $n=15$ (top panel) and for orbital parameters $\ha=0.9$, $\sigma=-0.5$, $p=12$, $e=0.8$ and $m=2$, $n=4$ (bottom panel). The red dots indicate the values at which the function is calculated during the numerical integration.}
  \label{fig:omega_t-m_phi}
\end{figure}

To calculate the energy and angular momentum fluxes and the strain at infinity, one has to find the partial amplitudes $\hat{C}^\pm_{lmn}$ and Eq.~\eqref{eq:Cpm_lmn_final} implies  integration over $\chi$. The numerical integration errors depend on the employed integration method and the number of points at which the function is enumerated. For our purposes, a fractional accuracy of the order of $10^{-6}$ is sufficient. Therefore, we used the midpoint rule inducing an error of the order $\mathcal{O}(N^{-2})$ to the integration, where $N$ is the number of points. The advantage of the midpoint rule is that for given accuracy, this method minimizes the number of points $N$ needed for the calculation. However, more complex method can be implemented in the future to improve the accuracy of this integration. The main oscillatory part of Eq.~\eqref{eq:Cpm_lmn_final} is contained in the exponential term $\exp(i D_r (\omega_{mn}t(\chi) - m\phi(\chi)))$. Figure~\ref{fig:omega_t-m_phi} shows the behavior of this oscillatory part for certain setups. The higher the value of $n$ is, the more the exponential function oscillates. High frequency oscillations are present especially around $\chi=\pi$ in high eccentricity cases. The number of the points $N$ needed for the integration is calculated from the maximum of the derivative of the function $\omega_{mn}t(\chi) - m\phi(\chi)$ with respect to $\chi$, which in dimensionless quantities reads $(\hat{\omega}_{mn} V^t(\hr(\chi)) - mV^\phi(\hr(\chi)))\sqrt{(1-e^2)/J(\chi)}/p$. 

The radial functions $R^\pm_{lmn}$ were calculated using the BHPToolkit \citep{BHPToolkit}, which employs the Mano-Suzuki-Takasugi (MST) method \citep{Sasaki:2003xr} or a numerical integration of the radial TE. The angular functions $S^{\ha\hat{\omega}}_{lm}$ were also calculated using the BHPToolkit which employs the Leaver's method \citep{Leaver:1985ax}.

\begin{figure}[!ht]
  \centering  
  \includegraphics[width=0.48\textwidth]{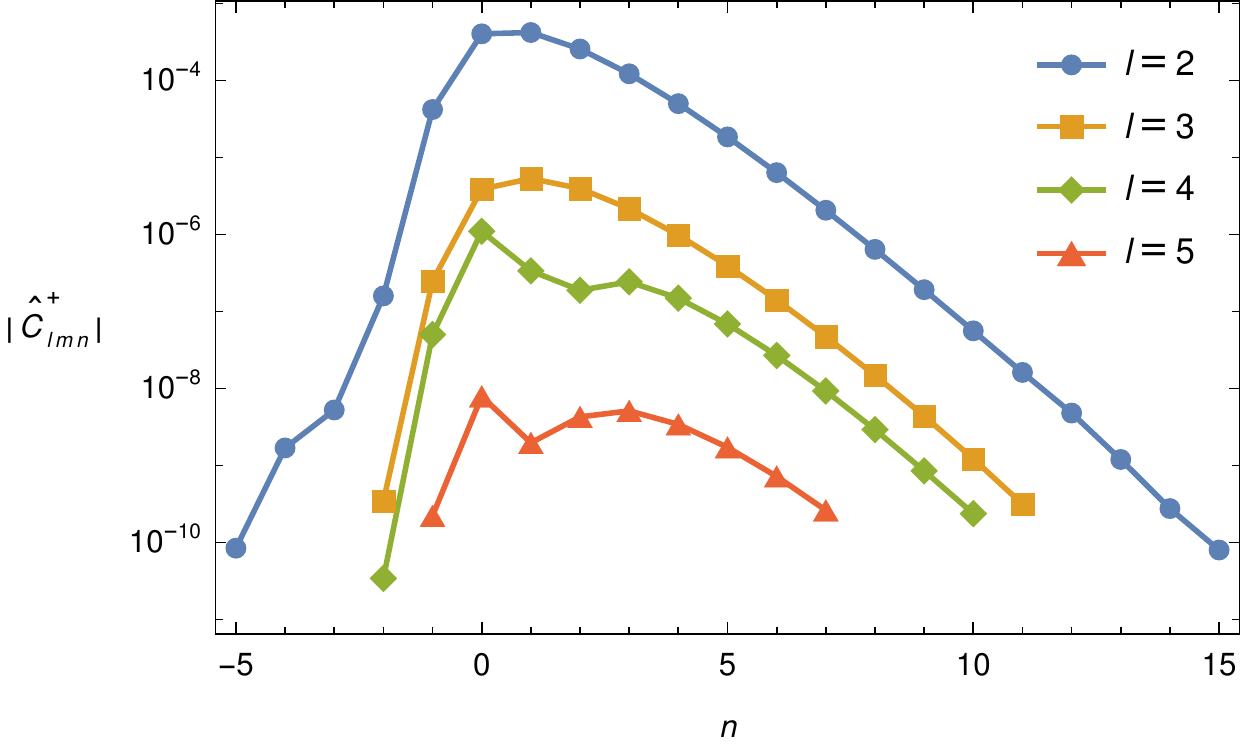} 
  \caption{Absolute values of the partial amplitudes $\abs{\hat{C}^+_{lmn}}$ for orbital parameters $\ha=0.9$, $\sigma=-0.5$, $p=12$, $e=0.2$ and azimuthal number $m=2$}
  \label{fig:modes_n}
\end{figure}

The strain is calculated from Eq. \eqref{eq:strain} and the fluxes are calculated from Eqs. \eqref{eq:fluxes}. The range of $l$ and $n$ for given $m$-mode was found in the following way. First we calculate the coefficient $\hat{C}^+_{lmn}$ for $l=\max(\abs{m},2)$ for a range of $n$ to find the mode with the maximal $\abs{\hat{C}^+_{lmn}}$. Then, we calculate other $l$ and $n$ modes until the absolute value is less than a chosen accuracy times the maximal mode. In our calculations, we have chosen accuracy $10^{-6}$. However, in some cases the absolute value $\abs{\hat{C}^+_{lmn}}$ is not monotonous in $n$ and it drops suddenly for some $n$. Because of this, after such a sudden decrease, amplitudes for more $n$ must be calculated. In Fig.~\ref{fig:modes_n}, the absolute values of the coefficients  $\abs{\hat{C}^+_{lmn}}$ are plotted for an orbit with $\ha=0.9$, $\sigma=-0.5$, $p=12$, $e=0.2$ and azimuthal number $m=2$ for different $l$ and $n$.  We can see that, for given accuracy, only limited number of modes is needed (for $l=m=2$ it is 21 $n$-modes) and the absolute value of the amplitudes is decreasing exponentially with $\abs{n}$ for sufficiently high $\abs{n}$. Note that although the astrophysical relevant value of the spin $\sigma$ is of the same order as the mass ratio $q\ll 1$, it is possible to calculate the GW fluxes for higher spins and then linearize the result in $\sigma$ to find the contribution of spin $\sigma\ll 1$. We use also these large values to make any deficiencies in our calculations prominent. 

In Appendix~\ref{sec:comaparisonDrasco} we compare our coefficients $\hat{C}^\pm_{lmn}$ and fluxes $\mathcal{F}^{E\infty}_{lmn}$ and $\mathcal{F}^{E{\rm H}}_{lmn}$ with that of \citep{Drasco:2005kz}. A simplified version of our code calculating GW fluxes from circular equatorial orbit of a spinning particle around a Kerr BH was used to independently verify the results of \citep{Piovano:2020}. These results are discussed in detail in \citep{Skoupy:2021}. Tables of the values of the partial amplitudes $\hat{C}^\pm_{lmn}$ for future references are in Appendix \ref{sec:tables}.

\subsubsection{Comparison of frequency domain and time domain}

To compare the time domain and the frequency domain results, we have calculated the coefficients $\hat{C}^+_{lmn}$ for some range of $l$ and $n$ in the frequency domain for different values of the spin $\sigma$ and of the eccentricity $e$. We have used these coefficients to find the respective strains and energy fluxes at infinity. Then, these results have served as reference values in our comparison with the azimuthal $m$-mode of the strain at infinity multiplied by the radial coordinate $\hr \, h_m$ and the energy fluxes at infinity $\mathcal{F}^{E \infty}_m$ obtained in the time domain. Because of the fact that the space discretization applied in \texttt{Teukode} induces numerical errors to the time domain calculations, we have run the time domain calculations for several resolutions and tested the convergence of the code.

To calculate the strains and the fluxes with in the time domain with \texttt{Teukode}, we need to approximate the delta functions representing the secondary body in the $\rho$ and $\theta$ directions. To do that we have used different combinations of Gaussian functions and piecewise polynomials in these directions. The accuracy appears to be higher when the piecewise polynomial are used in both $\rho$ and $\theta$ direction or Gaussian function in $\rho$ direction and piecewise polynomial in $\theta$ direction, than in the other two possible settings, i.e., Gaussian in both directions and Gaussian in $\theta$ direction with piecewise polynomial in $\rho$ direction. When the piecewise polynomial is used in both directions, calculations are faster and, therefore, we have used this approximation in most cases. In our calculations, the strain has been extracted at $r=\infty$ and $\theta=\pi/2$ and the energy flux has been averaged over two periods $T_r$ starting at the retarded coordinate around $u=350M$, where $u = t-r^\ast$.

\begin{figure}[!ht]
  \centering  
  \includegraphics[width=0.48\textwidth]{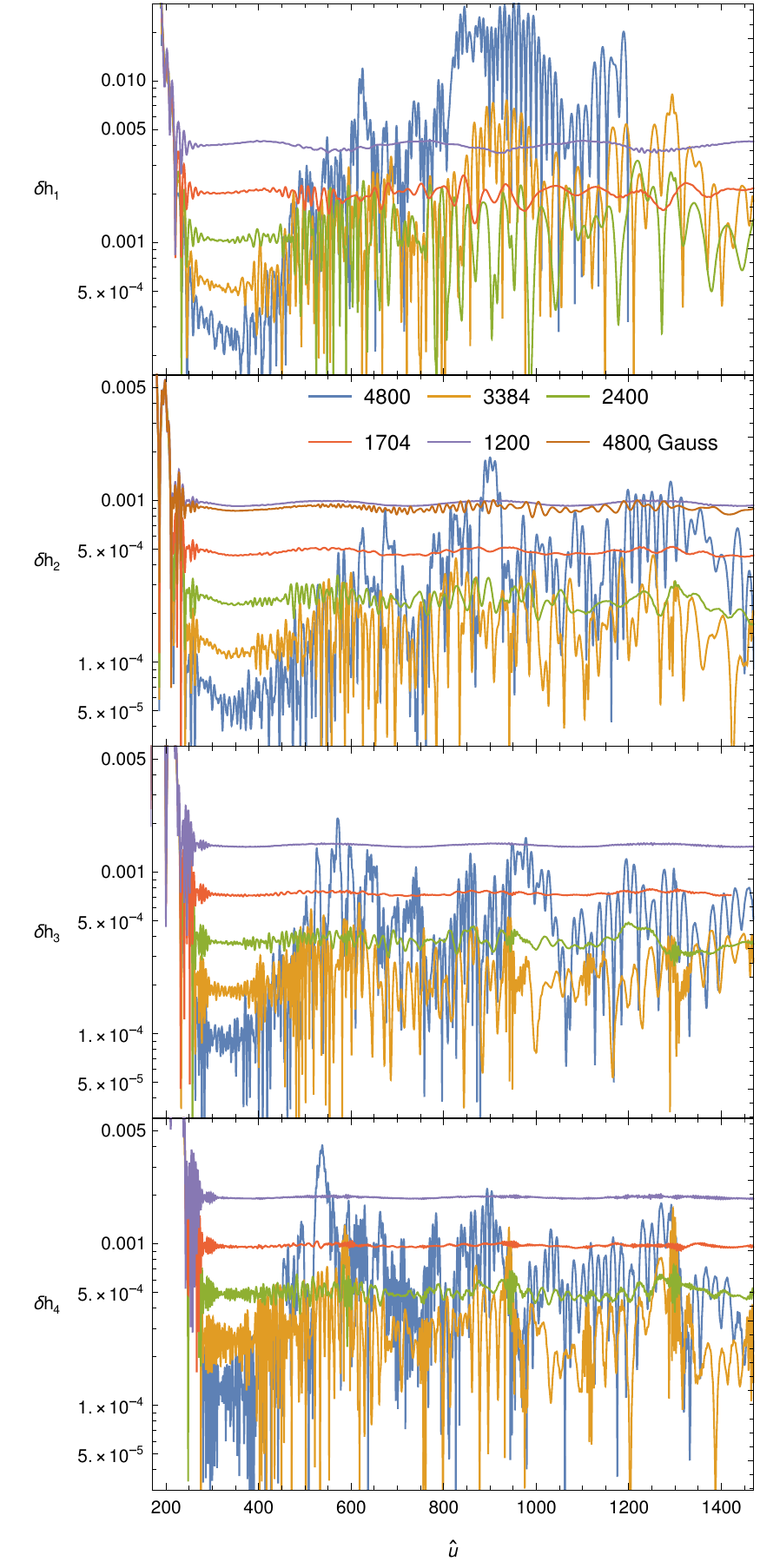} 
  \caption{The relative difference of the strain $\delta h_m$ from the $m=1$ mode (top panel) to the $m=4$ mode (bottom panel) as a function of the retarded coordinate $\hat{u}$ at $r=\infty$ and $\theta=\pi/2$. Each plotted curve represents a case with different number of points in the $\rho$ direction $N_\rho$. The piecewise polynomial approximation of the delta function was used for all cases apart from one, for which the Gaussian approximation with resolution $4800$ was employed. The parameters of the orbit are $\ha=0.9$, $\sigma=-0.5$, $p=12$, $e=0.2$. The initial noise is caused by zero initial data in time domain.}
  \label{fig:convergence_h}
\end{figure}

In order to provide a first comparison of the frequency and the time domain results, we use the relative difference of the azimuthal mode $m$ of the strain at $r=\infty$ and $\theta=\pi/2$
\begin{equation}\label{eq:deltah}
    \delta h_m = \abs{ 1-\frac{h_m^{\rm td}}{h_m^{\rm fd}} } \, ,
\end{equation}
where $h_m^{\rm td}$ is the strain calculated using \texttt{Teukode} and $h_m^{\rm fd}$ is $m$-mode of the strain calculated in frequency domain using Eq. \eqref{eq:strain} without the sum over $m$. Figure~\ref{fig:convergence_h} shows the relative difference of the azimuthal modes $m=1,2,3,4$ of the strain as function of the retarded coordinate $\hat{u}$. In this plot, the strain calculated in the frequency domain (the denominator of $\delta h_m$) remains fixed, while each time domain calculated evolution of the strain is performed for different number of points in the $\rho$ direction $N_\rho$ (resolution). The delta function is approximated by a piecewise polynomial for five resolutions ($N_\rho$=1200,1704,2400,3384,4800), while in one case is approximated by a Gaussian function for $N_\rho=4800$. We can see that the relative difference $\delta h_m$ tends to decrease as the resolution increases, but for the highest resolution $N_\rho=4800$ the numerical noise becomes significant. Though the Gaussian approximation is less accurate, the amplitude of its noise is relatively smaller than the amplitude of the noise for the piecewise polynomial approximation with the same resolution. We speculate that the cause of this numerical noise comes from the fact that as the resolution increases, the approximation becomes less smooth. Namely, we have used a 12th order approximation of the delta function, which is 12 points wide, for each resolution; therefore, the higher the resolution is, the narrower and higher is the delta function. Note that the $m=1$-mode has very small value and the noise has relatively higher amplitude than in $m=2,3,4$ modes. The $m=0$-mode, which is not shown here, although nonzero, has extremely small value allowing the numerical noise to be dominant.

\begin{figure}[!ht]
  \centering  
  \includegraphics[width=0.48\textwidth]{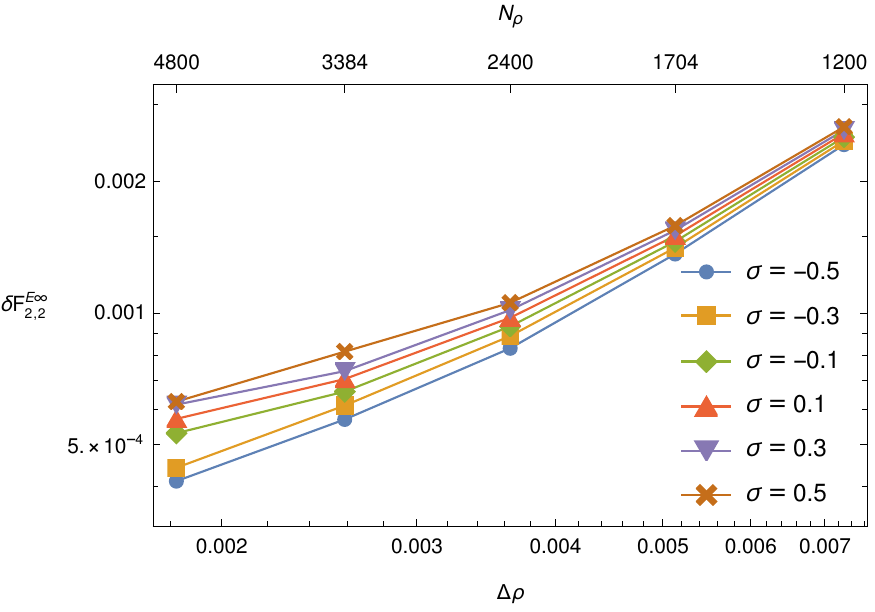}
  \caption{The relative difference of the energy flux $\delta \mathcal{F}^{E\infty}_{lm}$ of the $l=m=2$ mode as function of the grid length in the $\rho$ direction of the time domain calculations. Note that the time domain calculations have been projected on the $Y_{lm}$ basis, while the frequency domain ones on the $S_{lm}^{a\omega}$. Each curve represents a different value of the secondary spin, while the Kerr parameter $\ha=0.9$, semi-latus rectum $p=12$ and eccentricity $e=0.2$ remain fixed.}
  \label{fig:convergence2}
\end{figure}

To further check our results, we have calculated the relative difference of the energy fluxes
\begin{equation} \label{eq:deltaF}
    \delta \mathcal{F}^{E\infty}_{lm} = \abs{ 1-\frac{\mathcal{F}^{E\infty}_{lm,{\rm td}}}{\mathcal{F}^{E\infty}_{lm,{\rm fd}}} } \, ,
\end{equation}
where $\mathcal{F}^{E\infty}_{lm,{\rm td}}$ is the value calculated using \texttt{Teukode} and $\mathcal{F}^{E\infty}_{lm,{\rm fd}}$ is the value calculated with the frequency domain approach summed over $n$. Figure~\ref{fig:convergence2} shows how the time domain calculations of the dominant $l=m=2$ mode of the energy fluxes converges to the frequency ones as the resolution increases. For this plot we have kept fixed the Kerr parameter $\ha=0.9$, the semi-latus rectum $p=12$ and the eccentricity $e=0.2$, while we have used for each curve a different value of the secondary spin $\sigma$ spanning from $-0.5$ to $0.5$.
The relative difference in the fluxes should converge to zero as the grid length $\Delta\rho = (\rho_S - \rho_+)/N_\rho$ decreases  (increasing resolution). However, the relative differences do not converge to zero, because in the frequency domain calculations we use the projection to spin-weighted spheroidal harmonics $S_{lm}^{a\omega}$ and \texttt{Teukode} projects the strain to the spin-weighted spherical harmonics $Y_{lm} = S_{lm}^{0}$. For the dominant mode the difference between the projections to these functions is low because for low $a\omega$, the spheroidal functions $S_{lm}^{a\omega}$ can be approximated by the spherical functions $Y_{lm}$.

\begin{figure}[!ht]
  \centering  
  \includegraphics[width=0.48\textwidth]{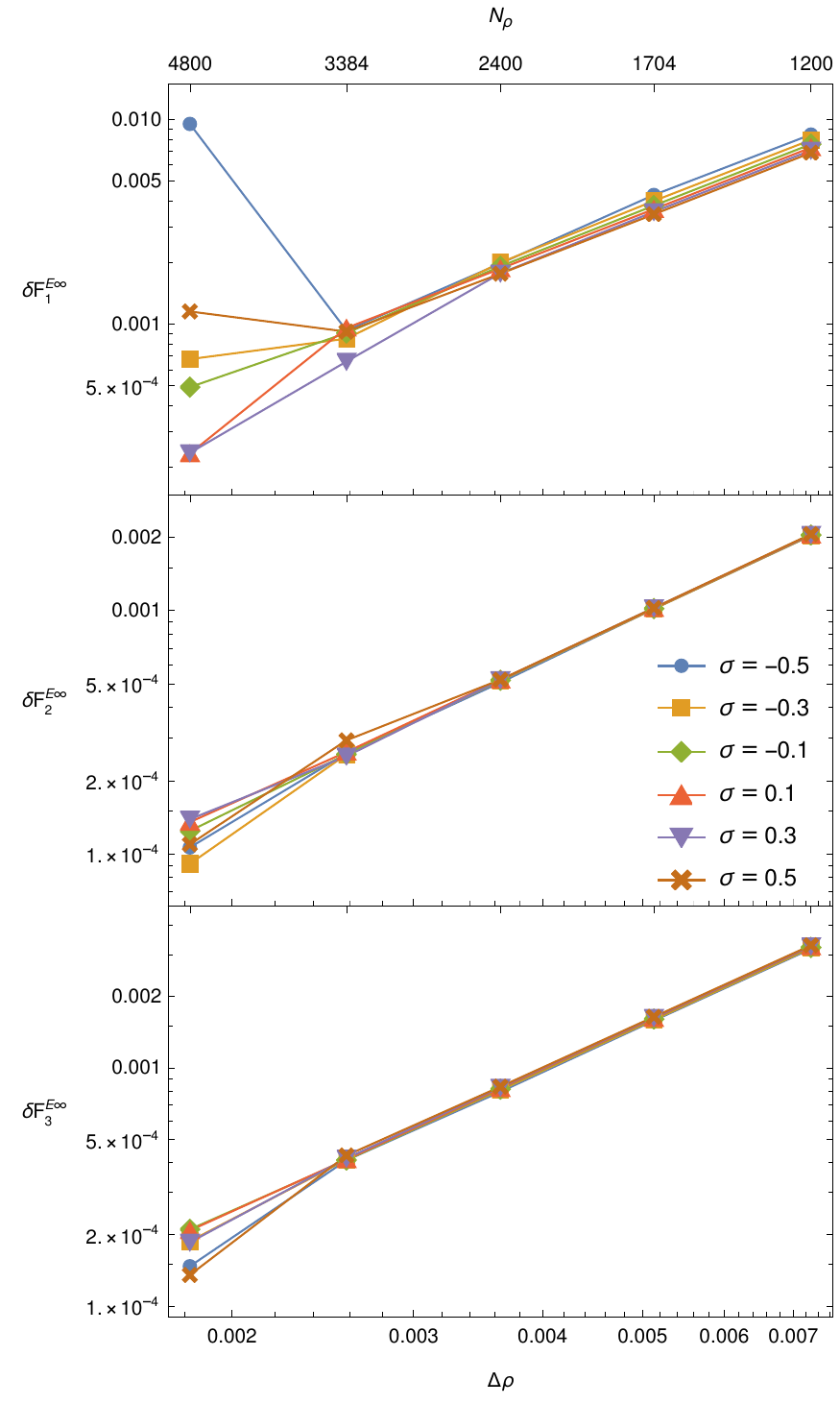}
  \caption{Comparison of frequency domain and time domain results. The relative difference $\delta \mathcal{F}^{E\infty}_{m=1}$ (top panel), $\delta\mathcal{F}^{E\infty}_{m=2}$ (middle panel) and $\delta \mathcal{F}^{E\infty}_{m=3}$ (bottom panel) is plotted for different values of the secondary spins $\sigma$ spanning from $-0.5$ to $0.5$. The Kerr parameter $\ha=0.9$, the semi-latus rectum $p=12$ and the eccentricity $0.2$ are kept fixed for all the cases.}
  \label{fig:convergence}
\end{figure}

Because of the aforementioned projection issue, for a proper comparison of the time and frequency domain results, we must calculate the sum of the fluxes over $l$. The relative difference
\begin{equation} \label{eq:deltaFm}
    \delta \mathcal{F}^{E\infty}_{m} = \abs{ 1-\frac{\mathcal{F}^{E\infty}_{m,{\rm td}}}{\mathcal{F}^{E\infty}_{m,{\rm fd}}} } \, ,
\end{equation}
for $m=1,2,3$ has been calculated for different secondary spins $\sigma$ in the frequency domain and in time domain we used different resolutions ($N_\rho=1200, 1704, 2400, 3384, 4800$). We can see in Fig.~\ref{fig:convergence} that the relative differences converge to zero as we expected. The lowest step $\Delta\rho$ corresponding to the highest resolution $N_\rho=4800$ shows variance in the relative differences. This is caused by the fact that the noise amplitude is the highest for the highest resolution, which can be seen in Fig.~\ref{fig:convergence_h}. Especially in the case $m=1$ where the energy flux is significantly lower than for $m=2$, the variance in the relative differences is clearly visible. For the highest resolution, the relative difference is higher for higher values of spin $\abs{\sigma}$. This can be caused by the numerical noise in time domain calculations induced by the non-smoothness of the piecewise polynomial approximation of the third derivative of the delta function. Namely, the term with the third derivative is proportional to the spin $\sigma$. For negative $\sigma$ the noise is relatively higher because the value of energy flux for $\sigma<0$ is lower than the flux for $\sigma>0$ and thus the noise is more dominant.

\begin{figure}[!ht]
  \centering  
  \includegraphics[width=0.48\textwidth]{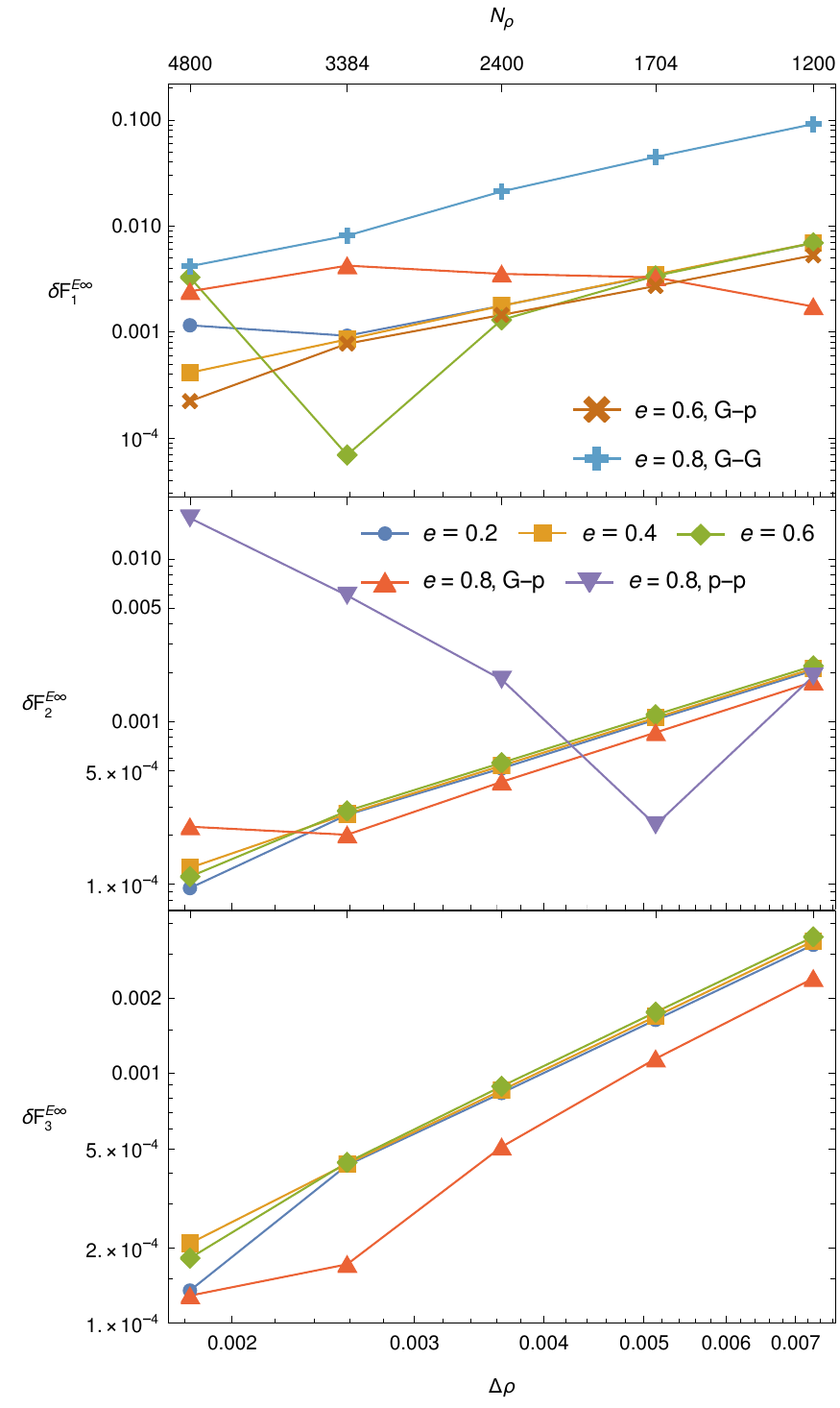}
  \caption{Comparison of frequency domain and time domain results. The relative difference $\delta \mathcal{F}^{E\infty}_{m=1}$ (top panel), $\delta\mathcal{F}^{E\infty}_{m=2}$ (middle panel) and $\delta \mathcal{F}^{E\infty}_{m=3}$ (bottom panel) is plotted for different values of the eccentricity $e$ spanning from $0.2$ to $0.8$. The Kerr parameter $\ha=0.9$, the secondary spin $\sigma=0.5$ and the semi-latus rectum $p=12$ are kept fixed for all the cases. If not specified, the delta function is approximated by a piecewise polynomial in both $\rho$ and $\theta$ direction. For $m=1$, $e=0.6$ and $m=2$, $e=0.8$ the delta function is approximated as Gaussian function in $\rho$ direction and piecewise polynomial in $\theta$ direction. For $m=1$, $e=0.8$ the delta function is approximated as Gaussian function in both $\rho$ and $\theta$ directions.}
  \label{fig:convergence_e}
\end{figure}

To check the dependence of our calculations on the value of eccentricity, we have calculated the energy fluxes for fixed Kerr parameter $\ha=0.9$, secondary spin $\sigma=0.5$ and semi-latus rectum $p=12$, while the eccentricity $e$ value spans from $0.2$ to $0.8$. For each eccentricity we have calculated the relative difference in the energy fluxes $\delta \mathcal{F}^{E\infty}_{m}$ for $m=1,2,3$. Then, we have compared the dependence of the relative difference on the resolution for different eccentricities as in the case with the changing secondary spin. This comparison is shown in Fig.~\ref{fig:convergence_e}. First, we have calculated the dominant $m=2$ mode in time domain with piecewise polynomial approximation of the delta function in both $\rho$ and $\theta$ direction (p-p), but for $e=0.8$ the noise is increasing with the resolution and $\delta\mathcal{F}^{E\infty}_2$ does not converge to zero (purple line in the middle panel of Fig.~\ref{fig:convergence_e}). Therefore, for $m=2$ and other modes, we performed the time domain calculations for $e=0.8$ using the Gaussian approximation in $\rho$ direction and the piecewise polynomial approximation in $\theta$ direction (G-p, red line in all panels of  Fig.~\ref{fig:convergence_e}). However, the $m=1$ mode has low amplitude and the noise is therefore more significant and the p-p approximation for $e=0.6$ and the G-p approximation for $e=0.8$ fails. Because of this, for $m=1$ mode we repeated the calculation for $e=0.6$ with G-p approximation and for $e=0.8$ with Gaussian approximation in both directions (G-G). 

The fact that for the piecewise polynomial approximation the noise has greater impact on higher eccentricities can be explained as follows. The shape of the delta function depends on the distance between the delta function and the two grid points around it. Since the distance between these grid points changes rapidly on a highly eccentric orbit, the shape of the delta function changes rapidly as well. The greater is the change in the shape, the greater is the noise. Thus, the piecewise polynomial approximation is optimal for circular trajectories.  Moreover, higher eccentricities imply longer periods of motion and thus longer runtime, which allows the exponentially growing noise to reach higher values. For the Gaussian approximation noise grows more slowly.

Figure~\ref{fig:convergence_e} indicates that by choosing a proper delta function approximation the relative difference $\delta\mathcal{F}^{E\infty}_m$ would converge to zero for all $m$-modes and eccentricities $e$. This choice currently seems to depend on the orbital parameters and modes.
For example, the piecewise polynomial approximation appears to be in general more efficient than the Gaussian approximation, however its own limitation in our example became prominent for high eccentricities and modes corresponding to small flux or strain absolute values, i.e. in modes that the numerical noise is dominant.

\section{Summary} \label{sec:conc}

In this work, we have studied the motion of a spinning particle in the equatorial plane of a Kerr black hole and the GW fluxes from these orbits. The only possible configuration of the spins in this setup is the spins to be parallel or antiparallel. In this framework, we have derived a reduced set of equations of motion equivalent to the MPD equations with TD SSC. Taking advantage of the fact that an orbit can be characterized by its constants of motion, namely the energy $E$ and the $z$ component of the total angular momentum $J_z$, we have provided explicit formulas for the energy and the angular momentum in terms of the eccentricity $e$ and semi-latus rectum $p$. Furthermore, through the reduced equations of motion and by introducing a Mino-like time parameter $\lambda$, we were able to find expressions allowing the numerical calculation of the frequencies of the radial and azimuthal motion. These expressions provide the frequencies with respect to $\lambda$ or the BL time. 

The orbital findings were then implemented in the calculation of the GW fluxes from the equatorial orbits in the frequency domain. Namely, this work introduces the formulas giving the strain $h$, the energy fluxes and the angular momentum fluxes at infinity and at the horizon from a spinning secondary moving on the equatorial plane of a Kerr black hole. For this purpose, we have developed a \textit{Mathematica} code calculating the amplitudes $C^\pm_{lmn}$ on which the frequency domain GW fluxes depend. We plan to make this code publicly available through the \textit{Black Hole Perturbation Toolkit} repository. The frequency domain results were, then, compared with time domain results obtained from a TE solver called \texttt{Teukode}. To improve the efficiency of \texttt{Teukode}, we have implemented a piecewise polynomial to approximate the delta functions and its derivatives in the spinning-particle source term. The comparison has shown good agreement between the frequency domain results with the time domain ones.

To check the discretization error in the time domain calculations introduced by the piecewise polynomial, we have calculated the fluxes in time domain for different resolutions and compared them with the respective frequency domain results. The difference between the results from these two approaches tend to consistently decrease with increasing resolution. However, for the highest resolution, which we have implemented, the numerical noise in the time domain calculations becomes significant. This behavior occurs for different calculation setups. Namely, we have checked our calculations by varying the secondary spin while keeping the other parameters fixed and  by varying the eccentricities while keeping the other parameters fixed.

These calculations are part of the on-going effort to build post-adiabatic gravitational waveforms  modelling gravitational waves emitted by extreme mass ratio inspirals. In a future work, the frequency domain fluxes will be used to find the adiabatic evolution of the orbit on the equatorial plane under the influence of radiation reaction. The influence of the secondary spin on the change of the orbital parameters and phase of the GW will be studied.


\begin{acknowledgments}
The authors have been supported by the fellowship Lumina Quaeruntur No. LQ100032102 of the Czech Academy of Sciences. The authors would like to acknowledge networking support by the GWverse COST Action CA16104, “Black holes, gravitational waves and fundamental physics''. V.S. would also like to express gratitude for the hospitality of the Theoretical Physics Institute at the University of Jena. We would like to thank Sebastiano Bernuzzi, Enno Harms, Vojt\v{e}ch Witzany and Tom\'{a}\v{s} Ledvinka for useful discussions and comments. This work makes use of the Black Hole Perturbation Toolkit. Computational resources were supplied by the project "e-Infrastruktura CZ" (e-INFRA LM2018140) provided within the program Projects of Large Research, Development and Innovations Infrastructures.
\end{acknowledgments}

\appendix

\section{List of dimensionless quantities}
\label{sec:quantities}

Throughout this work, we use several quantities both in their full form and dimensionless form. The dimensionless form is denoted by a hat. Their list with relation between the full and dimensionless form is in Table \ref{tab:dimensionless}. Some quantities such as the time parameter $\lambda$ or $x$ are defined only as dimensionless whereas other quantities are used only in their full form.

\begin{table}[!htb]
    \centering
    \caption{List of dimensionless quantities}
    \begin{tabular}{r@{$\,=\,$}l|l}
        $\hat{t}$ & $t/M$ & BL time \\
        $\hr$ & $r/M$ & BL radial coordinate \\
        $\ha$ & $a/M$ & Kerr parameter \\
        $\sigma$ & $S/(\mu M)$ & Secondary spin \\
        $\hat{E}$ & $E/\mu$ & Energy \\
        $\hat{J}_z$ & $J_z/(\mu M)$ & Angular momentum \\
        $\hat{\tau}$ & $\tau/M$ & Proper time \\
        $\hat{\Delta}$ & $\Delta/M^2$ & \\
        $\hat{\varpi}^2$ & $\varpi^2/M^2$ & \\
        $\hat{\Omega}_r$ & $\Omega_r M$ & Radial BL frequency \\
        $\hat{\Omega}_\phi$ & $\Omega_\phi M$ & Orbital BL frequency \\
        $\hat{\omega}$ & $\omega M$ & Frequency \\
        $\hat{C}_{ab}^0$ & $C_{ab}^0/\mu$ & \\
        $\hat{C}_{ab}^\sigma$ & $C_{ab}^\sigma/\mu$ & \\
        $\hat{C}^\pm_{lmn}$ & $C^\pm_{lmn}M^2/\mu$ & Partial amplitudes \\
        $\hat{u}$ & $u/M$ & Retarded coordinate
    \end{tabular}
    \label{tab:dimensionless}
\end{table}

\section{Formulas for GW fluxes}
\label{sec:coefficientsAi}

In this Appendix we derive the coefficients $A_i=A_i(r,\theta)$ and $B_{i+1}=B_{i+1}(r,\theta)$, $i=0,1,2$, in Eq. \eqref{eq:Ipm} for calculation of partial amplitudes of GWs from general bound orbits of a spinning particle around a Kerr black hole. Then we list explicit formulas for equatorial orbits with secondary spin parallel to the $z$ axis.

To find the form of the coefficients $A_i$ and $B_{i+1}$ in Eq.~\eqref{eq:Ipm}, the integrals \eqref{eq:source_Tlmomega} and \eqref{eq:Cpm_lmomega} must be evaluated using rules for integrating delta functions. We can classify the parts of the coefficients $A_i$ according to term from which they originate:
\begin{align}
    A_0 &= \sum_{ab=nn,n\mbar,\mbar\mbar} (A_{ab0}^0 + A_{ab0}^{t\phi} + A_{ab0}^{r} + A_{ab0}^{\theta}) \; , \\
    A_1 &= \sum_{ab=n\mbar,\mbar\mbar} (A_{ab1}^0 + A_{ab1}^{t\phi} + A_{ab1}^{r} + A_{ab1}^{\theta}) \; , \\
    A_2 &= A_{\mbar\mbar 0}^0 + A_{\mbar\mbar 0}^{t\phi} + A_{\mbar\mbar 1}^r + A_{\mbar\mbar 1}^{\theta} \; .
\end{align}
The terms $A_{abi}^0$ originate from the first term of the stress-energy tensor~\eqref{eq:T_ab} containing the nonspinning part of $T^{\mu\nu}$ and parts containing Christoffel symbols. The terms $A_{abi}^{t\phi}$ originate from the second term of \eqref{eq:T_ab} containing $t$ and $\phi$ derivative. Similarly, the terms $A_{abi}^r$ or $A_{abi}^\theta$ originate from the second term of Eq.~\eqref{eq:T_ab} containing $r$ or $\theta$ derivative respectively. The subscripts $ab$ denote the tetrad legs in Eq.~\eqref{eq:source_Tnm}.

$A_{abi}^0$ can be found by integrating $\theta$ and $\phi$ after substituting the first term of Eq.~\eqref{eq:T_ab} into Eq.~\eqref{eq:source_Tlmomega} by replacing $\theta\rightarrow\theta_p(t)$, $\phi\rightarrow\phi_p(t)$ and then using integration by parts in Eq.~\eqref{eq:Cpm_lmomega}, where the derivatives with respect to $r$ in \eqref{eq:source_Tnm} are shifted to the radial function $R^{\pm}_{lmn}$ to obtain
\begin{equation}
    A_{abi}^{0} = \left( C_{ab}^{0} - C_{ab}^{\sigma} \right) f_{ab}^{(i)} \; ,
\end{equation}
where $C^0_{ab}$ and $C^\sigma_{ab}$ are defined in Eq.~\eqref{eq:Cab}.

To find the form of $A_{abi}^{t\phi}$, we must perform integration by parts in Eq.~\eqref{eq:source_Tlmomega} where the $t$ or $\phi$ derivative in the second term of Eq.~\eqref{eq:T_ab} are shifted to $\exp(i\omega-im\phi)$ because no other functions depend on $t$ and $\phi$. From this, we get terms multiplied by $i\omega$ and $-im\phi$. After that, an integration over $r$ of Eq.~\eqref{eq:Cpm_lmomega} is done similarly as in the previous case and we obtain
\begin{equation}
    A_{abi}^{t\phi} = \frac{\rmd \tau}{\rmd t} (i \omega S^{t\mu} - i m S^{\phi\mu}) v^{\nu} e_{(\mu}^{(a)} e_{\nu)}^{(b)} f_{ab}^{(i)} \; .
\end{equation}

The term $A_{abi}^\theta$ is derived in similar way. The derivative with respect to $\theta$ in the second term in \eqref{eq:T_ab} is shifted to the functions $f_{ab}^{(i)}$ and the tetrad legs. The boundary term vanishes because $f_{ab}^{(i)}(r,0)=f_{ab}^{(i)}(r,\pi)=0$. The final term has the form
\begin{align}
    A_{abi}^\theta &= \frac{\rmd \tau}{\rmd t} S^{\theta(\mu} v^{\nu)} f_{ab}^{(i)} \partial_\theta (e_\mu^{(a)} e_{\nu}^{(b)}) \nonumber\\ & \;\;\;\;\; + \frac{\rmd \tau}{\rmd t} S^{\theta(\mu} v^{\nu)} e_{\mu}^{(a)} e_{\nu}^{(b)} \partial_\theta f_{ab}^{(i)} \; .
\end{align}

Now let us focus on the term containing the $r$ derivative in Eq.~\eqref{eq:T_ab}. After substituting the stress-energy tensor \eqref{eq:T_ab} into Eq.~\eqref{eq:source_Tnm}, the derivative of the delta function can be shifted to the function $f_{ab}^{(i)}$ and the tetrad legs. For example, from the first term of $\mathcal{T}_{n\mbar}$ we obtain
\begin{multline} \label{eq:exampleT_ab}
    \partial_r \left( f_{n\mbar}^{(1)}(r,\theta) n_\mu \mbar_\nu \partial_r \left( (v^t)^{-1} S^{r(\mu} v^{\nu)} \delta^3 \right) \right) = \\
    \partial_r^2 \left( f_{n\mbar}^{(1)}(r,\theta) n_\mu \mbar_\nu (v^t)^{-1} S^{r(\mu} v^{\nu)} \delta^3 \right) \\
    - \partial_r \left( \partial_r \left( f_{n\mbar}^{(1)}(r,\theta) n_\mu \mbar_\nu \right) (v^t)^{-1} S^{r(\mu} v^{\nu)} \delta^3 \right)
\end{multline}
After substituting Eq.~\eqref{eq:source_Tlmomega} into Eq.~\eqref{eq:Cpm_lmomega} we can change the order of the $t$ and $r$ integrals and integrate by parts. From the second term in Eq.~\eqref{eq:exampleT_ab} we obtain a term with derivatives with respect to $r$ of $f_{ab}^{(i)}$ and the tetrad legs
\begin{align}
    A_{abi}^r &= \frac{\rmd \tau}{\rmd t} S^{r(\mu} v^{\nu)} f_{ab}^{(i)} \partial_r (e_\mu^{(a)} e_{\nu}^{(b)}) \nonumber\\ & \;\;\;\;\; + \frac{\rmd \tau}{\rmd t} S^{r(\mu} v^{\nu)} e_{\mu}^{(a)} e_{\nu}^{(b)} \partial_r f_{ab}^{(i)} \; .
\end{align}
From the second term in Eq.~\eqref{eq:exampleT_ab} we obtain terms with one order higher derivatives of the radial function $R^\pm_{lm\omega}$, of which the integration by parts we can perform to obtain the coefficients
\begin{align}
    B_1 &= \sum_{ab=nn,n\mbar,\mbar\mbar} B_{ab1} \; , \\
    B_2 &= \sum_{ab=n\mbar,\mbar\mbar} B_{ab2} \; , \\
    B_3 &= B_{\mbar\mbar 3} \, ,
\end{align}
where
\begin{align}
    B_{ab(i+1)} &= -\frac{\rmd \tau}{\rmd t} S^{r(\mu} v^{\nu)} e_\mu^{(a)} e_{\nu}^{(b)} f_{ab}^{(i)} \; .
\end{align}

The functions $f^{(i)}_{ab} = f^{(i)}_{ab}(r,\theta)$ in the equatorial plane are given by
\begin{align}
    f_{nn}^{(0)}\left(r,\frac{\pi}{2}\right) &= -\frac{2r^2}{\Delta^2} \left( \mathscr{L}^\dag_1 \mathscr{L}^\dag_2 - \frac{2i a}{r} \mathscr{L}^\dag_2 \right) S_{lm}^{a\omega}\left(\theta\right) \Bigr|_{\theta \rightarrow \frac{\pi}{2}} \; , \\
    f_{n\mbar}^{(0)}\left(r,\frac{\pi}{2}\right) &= \frac{2\sqrt{2} r}{\Delta} \left( \frac{i K}{\Delta} +\frac{2}{r} \right) \mathscr{L}^\dag_2 S_{lm}^{a\omega}\left(\theta\right) \Bigr|_{\theta \rightarrow \frac{\pi}{2}} \; , \\
    f_{n\mbar}^{(1)}\left(r,\frac{\pi}{2}\right) &= \frac{2\sqrt{2} r}{\Delta} \mathscr{L}^\dag_2 S_{lm}^{a\omega}\left(\theta\right) \Bigr|_{\theta \rightarrow \frac{\pi}{2}} \; , \\
    f_{\mbar\mbar}^{(0)}\left(r,\frac{\pi}{2}\right) &= \left(i\partial_r \left( \frac{K}{\Delta} \right) -2i\frac{K}{\Delta r} + \frac{K^2}{\Delta^2} \right) S_{lm}^{a\omega}\left(\frac{\pi}{2}\right) \; , \\
    f_{\mbar\mbar}^{(1)}\left(r,\frac{\pi}{2}\right) &= -2 \left( \frac{1}{r} + i \frac{K}{\Delta} \right) S_{lm}^{a\omega}\left(\frac{\pi}{2}\right) \; , \\
    f_{\mbar\mbar}^{(2)}\left(r,\frac{\pi}{2}\right) &= -S_{lm}^{a\omega}\left(\frac{\pi}{2}\right)\, ,
\end{align}
where
\begin{align}
    K &= (r^2+a^2)\omega - am \; , \\
    \mathscr{L}^\dag_n &= \partial_\theta -m\csc\theta + a\omega \sin\theta + n\cot\theta \; .
\end{align}

Up to this point the analysis holds for generic orbits of a spinning particle. When we constrain the particle on equatorial orbits with its spin set parallel to the $z$ axis, then $S^{\theta\mu}=0$ for all $\mu$ and, therefore, $A^{\theta}_{abi}=0$. For the presentation of the equatorial case, we prefer to use the dimensionless quantities.

In the definition of $C^0_{ab}$ and $C^\sigma_{ab}$ \eqref{eq:Cab} we can replace the derivative with respect to $\tau$ in $v^\mu$ with derivative with respect to $\lambda$ and use the fact that $V^t = \rmd \hat{t}/\rmd \lambda$, $V^r = \rmd \hat{r}/\rmd \lambda$ and $V^\phi = \rmd \phi/\rmd \lambda$. From Eqs.~\eqref{eq:specific_linmom_CEO_TUL} and \eqref{eq:EOM_equatorial} we obtain
\begin{align}
    \hat{C}^0_{nn} &= \frac{\rmd \lambda}{\rmd \hat{t}}  \frac{V_n^2}{\Sigma_\sigma} \; , \\
    \hat{C}^0_{n\mbar} &= \frac{\rmd \lambda}{\rmd \hat{t}} \frac{V_{\mbar} V_n \left( 2 \hat{r}^3+\sigma^2 \right)}{2 \Sigma_\sigma \left(\hat{r}^3+2 \sigma^2\right)} \; , \\
    \hat{C}^0_{\mbar\mbar} &= \frac{\rmd \lambda}{\rmd \hat{t}} \frac{\hat{r} V_{\mbar}^2}{(\hat{r}^3+2 \sigma^2)} \; ,
\end{align}
\begin{widetext}
\begin{align}
    \hat{C}^\sigma_{nn} &= \frac{\rmd \lambda}{\rmd \hat{t}} \frac{\sigma}{2 \hr^2 \Sigma_\sigma} \left( 2 \ha V_n^2 - \frac{\hat{\Delta} (\hr^3+2\sigma^2)}{\hr^3-\sigma^2} V_n x + (\ha^2-\hr) V^r x \right) \; , \\
    \hat{C}^\sigma_{n\mbar} &= \frac{\rmd \lambda}{\rmd \hat{t}} \frac{i\sigma}{2\sqrt{2} \hr \Sigma_\sigma} \left( -\frac{\ha^2-\hr}{\Delta} \left( 2V_n^2 + \left(V^r\right)^2 \right) - \frac{\ha^2-\hr^2}{\hat{\Delta}} V_n V^r + \frac{3\ha\sigma^2}{\hr \Sigma_\sigma} V_n x - \ha V^r x + \frac{\hat{\Delta}(\hr^3+2\sigma^2)}{2\hr \Sigma_\sigma} x^2 \right) \; , \\
    \hat{C}^\sigma_{\mbar\mbar} &= \frac{\rmd \lambda}{\rmd \hat{t}} \frac{\sigma}{\Sigma_\sigma} \left( -\frac{\ha}{\hat{\Delta}} \left( 2V_n^2 + 2 V_n V^r + \left(V^r\right)^2 \right) + \frac{i}{\hr\sqrt{2}} \left( 2 V_n + V^r \right) V_{\mbar}  \right)\, ,
\end{align}
\end{widetext}
where
\begin{align}
    V_n &= V^t n_t + V^r n_r + V^\phi \frac{n_\phi}{M} = - \frac{P_\sigma(\hr) + V^r}{2} \; , \\
    V_{\mbar} &= V^t \mbar_t + V^\phi \frac{\mbar_\phi}{M} = -\frac{i x \left( \hr^3+2 \sigma^2 \right)}{\sqrt{2} \Sigma_\sigma} \; .
\end{align}

We can rewrite the expressions for $A_{abi}^{t\phi}$, $A_{abi}^r$, and $B_{ab(i+1)}$ into dimensionless quantities as
\begin{align}
    \hat{A}_{abi}^{t\phi} &= \frac{\rmd \lambda}{\rmd \hat{t}} \left( i\hat{\omega} \hat{S}^{t}{}_{(a} - im\hat{S}^{\phi}{}_{(a} \right) V_{b)} \hat{f}_{ab}^{(i)}\left(\hat{r},\frac{\pi}{2}\right) \; , \\
    \hat{A}_{abi}^r &= \frac{\rmd \lambda}{\rmd \hat{t}} (\hat{S}^{r}{}_{(\partial_{\hat{r}}a} V_{b)} + \hat{S}^{r}{}_{(b} V_{\partial_{\hat{r}}a)}) f_{ab}^{(i)}\left(\hat{r},\frac{\pi}{2}\right) \nonumber\\ & \;\;\;\;\; + \frac{\rmd \lambda}{\rmd \hat{t}} \hat{S}^{r}{}_{(a} V_{b)} \partial_{\hat{r}} \hat{f}_{ab}^{(i)}\left(\hat{r},\frac{\pi}{2}\right) \; , \\
    \hat{B}_{ab(i+1)} &= -\frac{\rmd \lambda}{\rmd \hat{t}} \hat{S}^{r}{}_{(a} V_{b)} \hat{f}_{ab}^{(i)}\left(\hat{r},\frac{\pi}{2}\right)\, ,
\end{align}
where we used the dimensionless projections of $S^{\mu\nu}$ into the tetrad
\begin{align}
    \hat{S}^{t}{}_{n} &= \frac{1}{\mu M} \left(S^{tr} n_r + S^{r\phi}  n_\phi \right) = \frac{\sigma \left(x \hat{\varpi}^2 - 2 \ha V_n \right)}{2 \hr \Sigma_\sigma} \; , \\
    \hat{S}^{r}{}_n &= \frac{1}{\mu M} \left(-S^{tr} n_t + S^{r\phi}  n_\phi \right) = -\frac{\sigma x \hat{\Delta}}{2 \hr \Sigma_\sigma} \; , \\
    \hat{S}^{\phi}{}_n &= \frac{1}{\mu} \left(-S^{t\phi} n_t - S^{r\phi} n_r \right) = \frac{\sigma (\ha x - 2V_n)}{2 \hr \Sigma_\sigma } \; , \\
    \hat{S}^{t}{}_{\mbar} &= \frac{1}{\mu M} S^{t\phi} \mbar_\phi = -\frac{i \sigma \hat{\varpi}^2 V^r}{\sqrt{2} \hat{\Delta} \Sigma_\sigma} \; , \\
    \hat{S}^{r}{}_{\mbar} &= \frac{1}{\mu M} \left(-S^{tr} \mbar_t + S^{r\phi} \mbar_\phi \right) = -\frac{i \sigma P_\sigma(\hr)}{\sqrt{2} \Sigma_\sigma} \; , \\
    \hat{S}^{\phi}{}_{\mbar} &= -\frac{1}{\mu} S^{t\phi} \mbar_t = -\frac{i \sigma \ha V^r}{\sqrt{2} \hat{\Delta} \Sigma_\sigma} \; .
\end{align}

The  quantities $V_{\partial_{\hat{r}} a}$ and $\hat{S}^r{}_{\partial_{\hat{r}} a}$ can be understood as dimensionless projections on the differentiated tetrad $\partial_r e^\mu_{(a)}$
\begin{align}
    V_{\partial_{\hat{r}} n} &= M \left( V^t \partial_r n_t + V^r \partial_r n_r + V^\phi \frac{\partial_r n_\phi}{M} \right) \nonumber \\ &= \frac{(\ha^2-\hr) P_\sigma(\hr)}{\hr \hat{\Delta}} \; , \\
    V_{\partial_{\hat{r}} \mbar} &= M \left( V^t \partial_r \mbar_t + V^\phi \frac{\partial_r \mbar_\phi}{M} \right) \nonumber \\ &= \frac{V_{\mbar}}{\hat{r}} - \frac{i\sqrt{2} \hat{a} P_\sigma(\hat{r})}{\hat{\Delta}} \; , \\
    \hat{S}^{r}{}_{\partial_{\hat{r}} n} &= \frac{1}{\mu} \left( -S^{tr} \partial_r n_t + S^{r\phi} \partial_r n_\phi \right) = \frac{\sigma (\ha^2-\hr) x}{\hr^2 \Sigma_\sigma} \; , \\
    \hat{S}^{r}{}_{\partial_{\hat{r}} \mbar} &= \frac{1}{\mu} \left( -S^{tr} \partial_r \mbar_t + S^{r\phi} \partial_r \mbar_\phi \right) \nonumber \\ &= -\frac{i \sigma (2 \ha x + P_\sigma(\hr))}{ \sqrt{2} \hr \Sigma_\sigma} \, ,
\end{align}
where the covariant components of the tetrad are
\begin{align}
    n_\mu &= \frac{1}{2\Sigma} \left( -\Delta, -\Sigma, 0, a \Delta \sin^2\theta \right) \; , \\
    \mbar_\mu &= -\frac{\rho}{\sqrt{2}} \left( i a \sin\theta, 0, \Sigma, -i \varpi^2 \sin\theta \right) \; .
\end{align}

\section{Comparison with [Phys. Rev. D 73, 024027 (2006)]}
\label{sec:comaparisonDrasco}

This section compares our frequency domain calculations for a nonspinning particle with results obtained in \cite{Drasco:2005kz}. In that work, the GW fluxes were calculated from generic orbits of a nonspinning particle moving around a Kerr black hole using Teukolsky formalism with the fractional accuracy of the energy flux $l, m$-modes set to $10^{-6}$. 

\begin{figure}[!ht]
  \centering  
  \includegraphics[width=0.48\textwidth]{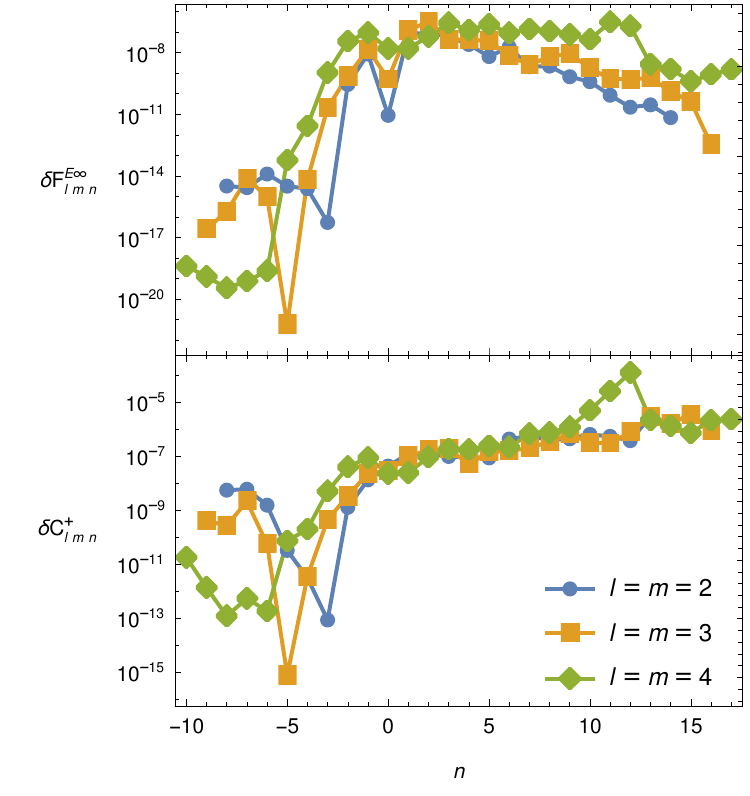}
  \caption{Differences between our frequency domain results and the results obtained in \citep{Drasco:2005kz}. Top panel: the difference $\delta \mathcal{F}^{E\infty}_{lmn}$ between the fluxes normalized by $\max\limits_{n_{\rm min}\leq n \leq n_{\rm max}}\abs{\mathcal{F}^{E\infty}_{lmn}}$. Bottom panel: the difference $\delta \hat{C}^+_{lmn}$ between the coefficients normalized by $\max\limits_{n_{\rm min}\leq n \leq n_{\rm max}}\abs{\hat{C}^{+}_{lmn}}$.}
  \label{fig:hughes}
\end{figure}

\begin{figure}[!ht]
  \centering  
  \includegraphics[width=0.48\textwidth]{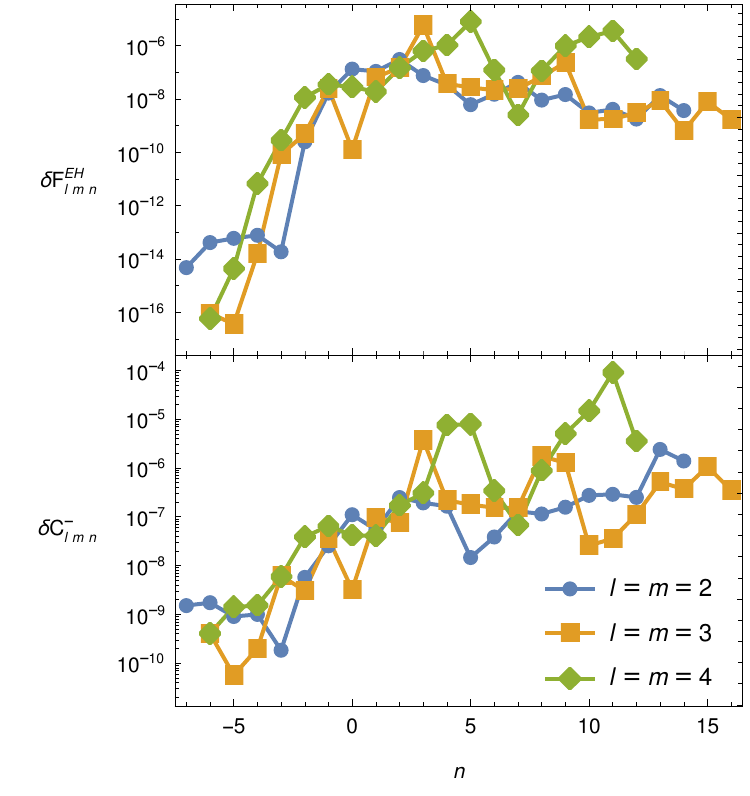}
  \caption{Differences between our frequency domain results and the results obtained in \citep{Drasco:2005kz}. Top panel: the difference $\delta \mathcal{F}^{E {\rm H}}_{lmn}$ between the fluxes normalized by $\max\limits_{n_{\rm min}\leq n \leq n_{\rm max}}\abs{\mathcal{F}^{E {\rm H}}_{lmn}}$. Bottom panel: the difference $\delta \hat{C}^-_{lmn}$ between the coefficients normalized by $\max\limits_{n_{\rm min}\leq n \leq n_{\rm max}}\abs{\hat{C}^{-}_{lmn}}$.}
  \label{fig:hughesH}
\end{figure}

We have compared our data with theirs for an equatorial orbit around a Kerr black hole with $\ha=0.3$, $p=8.463649=1.7\hr_{\rm ISCO}$ and $e=0.3$. In particular, we have compared our energy fluxes $\mathcal{F}^{E\infty}_{lmn}$, $\mathcal{F}^{E {\rm H}}_{lmn}$ and amplitudes $\hat{C}^\pm_{lmn}$ with their data. In the top panel of Fig.~\ref{fig:hughes}, we plot the difference between our calculated fluxes $\mathcal{F}^{E\infty}_{lmn}$ and the fluxes $\mathcal{F}^{E\infty}_{lmn{\rm DH}}$ calculated in \cite{Drasco:2005kz} normalized by the maximum of $\mathcal{F}^{E\infty}_{lmn}$ over $n$ for each $lm$-mode
\begin{equation}
    \delta \mathcal{F}^{E\infty}_{lmn} = \frac{\abs{\mathcal{F}^{E\infty}_{lmn}-\mathcal{F}^{E\infty}_{lmn{\rm DH}}}}{\max\limits_{n_{\rm min}\leq n \leq n_{\rm max}}\mathcal{F}^{E\infty}_{lmn}}\, .
\end{equation}
We can see that for each $lmn$-mode the error is less than $10^{-6}$ of the maximal value for given $l$ and $m$. In a similar way, we have compared the coefficients $\hat{C}^+_{lmn}$ using the quantity
\begin{equation}
    \delta C^+_{lmn} = \frac{\abs{\hat{C}^+_{lmn}-\hat{C}^+_{lmn{\rm DH}}}}{\max\limits_{n_{\rm min}\leq n \leq n_{\rm max}}\abs{\hat{C}^+_{lmn}}} \; .
\end{equation}
The result of this comparison is shown in the bottom panel of Fig.~\ref{fig:hughes}. The normalized difference for the coefficients $\hat{C}^+_{lmn}$ is higher than in the flux comparison, because the flux is calculated from the second power of $\hat{C}^+_{lmn}$ and the error is thus relatively smaller. Similar comparison was calculated for the horizon fluxes $\mathcal{F}^{E {\rm H}}_{lmn}$ and $\hat{C}^\pm_{lmn}$. The result is shown in Fig.~\ref{fig:hughesH}. Although the accuracy is less than $10^{-6}$ for some modes, the contribution from the horizon fluxes is smaller than from the fluxes to infinity and the overall accuracy remains higher.

\section{Data tables}
\label{sec:tables}

In this appendix we present data tables of the partial amplitudes $C^\pm_{lmn}$ (Tables \ref{tab:tab_m1} to \ref{tab:tab_m4}) for an orbit with orbital parameters $\ha=0.9$, $\sigma=-0.5$, $p=12$, $e=0.2$. The constants of motion and the fundamental frequencies calculated from the Eqs. \eqref{eq:energy2}, \eqref{eq:angmom2}, \eqref{eq:Omega_phi} and \eqref{eq:Omega_r} are
\begin{align*}
    \hat{E} &= 0.961918749642517680134729458401233368989\ldots \\
    \hat{J}_z &= 3.322244358788816670183960181110056686457\ldots \\
    \hat{\Omega}_\phi &= 0.022671787375747548523093927931917807 \\
    \hat{\Omega}_r &= 0.017744448092313388568850328609190010
\end{align*}
Only modes with $\abs{C^+_{lmn}}>10^{-9}$ are listed for $1\leq m\leq 4$. The accuracy of the dominant modes should be at six significant digits, but for lower modes, the accuracy drops. This accuracy depends mostly on the accuracy of the radial function $R^{\pm}_{lmn}$ and the coordinates $t(\chi)$ and $\phi(\chi)$.

 \clearpage
 \onecolumngrid

\begin{table}[htbp!]
    \centering
    \caption{List of partial amplitudes $C^{\pm}_{l1n}$ for an orbit with orbital parameters $\ha=0.9$, $\sigma=-0.5$, $p=12$, $e=0.2$.}
    \label{tab:tab_m1}
    \begin{tabular}{|r|r|r|r|r|r|r|} \hline
\multicolumn{1}{|c}{$l$}	&	\multicolumn{1}{|c}{$m$}	&	\multicolumn{1}{|c}{$n$}	& \multicolumn{1}{|c}{${\rm Re}\{C^+_{lmn}\}$}	&	\multicolumn{1}{|c}{${\rm Im}\{C^+_{lmn}\}$}	&	\multicolumn{1}{|c}{${\rm Re}\{C^-_{lmn}\}$}	&	\multicolumn{1}{|c|}{${\rm Im}\{C^-_{lmn}\}$}	\\ \hline
2	&	1	&	-6	&	5.167891$\times 10^{-10}$	&	-1.467715$\times 10^{-9\phantom{0}}$	&	4.763386$\times 10^{-9\phantom{0}}$	&	7.282975$\times 10^{-8\phantom{0}}$	\\ \hline
2	&	1	&	-5	&	1.335467$\times 10^{-9\phantom{0}}$	&	-4.271458$\times 10^{-9\phantom{0}}$	&	9.287132$\times 10^{-9\phantom{0}}$	&	3.018417$\times 10^{-7\phantom{0}}$	\\ \hline
2	&	1	&	-4	&	2.722364$\times 10^{-9\phantom{0}}$	&	-1.043309$\times 10^{-8\phantom{0}}$	&	-5.859804$\times 10^{-10}$	&	1.237186$\times 10^{-6\phantom{0}}$	\\ \hline
2	&	1	&	-3	&	3.539058$\times 10^{-9\phantom{0}}$	&	-1.824696$\times 10^{-8\phantom{0}}$	&	-1.435506$\times 10^{-7\phantom{0}}$	&	5.014982$\times 10^{-6\phantom{0}}$	\\ \hline
2	&	1	&	-2	&	1.292175$\times 10^{-9\phantom{0}}$	&	-1.244299$\times 10^{-8\phantom{0}}$	&	-1.084857$\times 10^{-6\phantom{0}}$	&	2.011327$\times 10^{-5\phantom{0}}$	\\ \hline
2	&	1	&	-1	&	2.025147$\times 10^{-10}$	&	4.098935$\times 10^{-9\phantom{0}}$	&	-6.110857$\times 10^{-6\phantom{0}}$	&	7.974254$\times 10^{-5\phantom{0}}$	\\ \hline
2	&	1	&	0	&	-8.063520$\times 10^{-7\phantom{0}}$	&	-5.100353$\times 10^{-6\phantom{0}}$	&	-4.159829$\times 10^{-5\phantom{0}}$	&	4.294392$\times 10^{-4\phantom{0}}$	\\ \hline
2	&	1	&	1	&	-1.456843$\times 10^{-6\phantom{0}}$	&	-6.201949$\times 10^{-6\phantom{0}}$	&	-3.035062$\times 10^{-5\phantom{0}}$	&	2.643207$\times 10^{-4\phantom{0}}$	\\ \hline
2	&	1	&	2	&	-1.100808$\times 10^{-6\phantom{0}}$	&	-3.735099$\times 10^{-6\phantom{0}}$	&	-1.503776$\times 10^{-5\phantom{0}}$	&	1.150093$\times 10^{-4\phantom{0}}$	\\ \hline
2	&	1	&	3	&	-5.586369$\times 10^{-7\phantom{0}}$	&	-1.633602$\times 10^{-6\phantom{0}}$	&	-6.182375$\times 10^{-6\phantom{0}}$	&	4.266183$\times 10^{-5\phantom{0}}$	\\ \hline
2	&	1	&	4	&	-2.230519$\times 10^{-7\phantom{0}}$	&	-5.891306$\times 10^{-7\phantom{0}}$	&	-2.272133$\times 10^{-6\phantom{0}}$	&	1.441687$\times 10^{-5\phantom{0}}$	\\ \hline
2	&	1	&	5	&	-7.504175$\times 10^{-8\phantom{0}}$	&	-1.849669$\times 10^{-7\phantom{0}}$	&	-7.750262$\times 10^{-7\phantom{0}}$	&	4.582582$\times 10^{-6\phantom{0}}$	\\ \hline
2	&	1	&	6	&	-2.190751$\times 10^{-8\phantom{0}}$	&	-5.168866$\times 10^{-8\phantom{0}}$	&	-2.508722$\times 10^{-7\phantom{0}}$	&	1.395109$\times 10^{-6\phantom{0}}$	\\ \hline
2	&	1	&	7	&	-5.569707$\times 10^{-9\phantom{0}}$	&	-1.283265$\times 10^{-8\phantom{0}}$	&	-7.819832$\times 10^{-8\phantom{0}}$	&	4.114092$\times 10^{-7\phantom{0}}$	\\ \hline
2	&	1	&	8	&	-1.192995$\times 10^{-9\phantom{0}}$	&	-2.742169$\times 10^{-9\phantom{0}}$	&	-2.371637$\times 10^{-8\phantom{0}}$	&	1.184135$\times 10^{-7\phantom{0}}$	\\ \hline
3	&	1	&	-6	&	4.142460$\times 10^{-10}$	&	-9.135246$\times 10^{-10}$	&	1.272727$\times 10^{-8\phantom{0}}$	&	1.107665$\times 10^{-8\phantom{0}}$	\\ \hline
3	&	1	&	-5	&	9.739033$\times 10^{-10}$	&	-2.513096$\times 10^{-9\phantom{0}}$	&	4.917854$\times 10^{-8\phantom{0}}$	&	4.243317$\times 10^{-8\phantom{0}}$	\\ \hline
3	&	1	&	-4	&	1.677799$\times 10^{-9\phantom{0}}$	&	-5.375419$\times 10^{-9\phantom{0}}$	&	1.880053$\times 10^{-7\phantom{0}}$	&	1.599580$\times 10^{-7\phantom{0}}$	\\ \hline
3	&	1	&	-3	&	1.428809$\times 10^{-9\phantom{0}}$	&	-6.372160$\times 10^{-9\phantom{0}}$	&	7.093426$\times 10^{-7\phantom{0}}$	&	5.919756$\times 10^{-7\phantom{0}}$	\\ \hline
3	&	1	&	-2	&	-3.968943$\times 10^{-10}$	&	3.427931$\times 10^{-9\phantom{0}}$	&	2.633559$\times 10^{-6\phantom{0}}$	&	2.144669$\times 10^{-6\phantom{0}}$	\\ \hline
3	&	1	&	0	&	-1.196768$\times 10^{-7\phantom{0}}$	&	-6.663590$\times 10^{-7\phantom{0}}$	&	4.292987$\times 10^{-5\phantom{0}}$	&	3.279176$\times 10^{-5\phantom{0}}$	\\ \hline
3	&	1	&	1	&	2.815396$\times 10^{-7\phantom{0}}$	&	1.022378$\times 10^{-6\phantom{0}}$	&	2.784472$\times 10^{-5\phantom{0}}$	&	2.044711$\times 10^{-5\phantom{0}}$	\\ \hline
3	&	1	&	2	&	3.393782$\times 10^{-7\phantom{0}}$	&	9.521505$\times 10^{-7\phantom{0}}$	&	1.277421$\times 10^{-5\phantom{0}}$	&	8.974092$\times 10^{-6\phantom{0}}$	\\ \hline
3	&	1	&	3	&	1.975341$\times 10^{-7\phantom{0}}$	&	4.622740$\times 10^{-7\phantom{0}}$	&	4.983533$\times 10^{-6\phantom{0}}$	&	3.333072$\times 10^{-6\phantom{0}}$	\\ \hline
3	&	1	&	4	&	8.151849$\times 10^{-8\phantom{0}}$	&	1.663691$\times 10^{-7\phantom{0}}$	&	1.766440$\times 10^{-6\phantom{0}}$	&	1.119246$\times 10^{-6\phantom{0}}$	\\ \hline
3	&	1	&	5	&	2.659713$\times 10^{-8\phantom{0}}$	&	4.876757$\times 10^{-8\phantom{0}}$	&	5.874042$\times 10^{-7\phantom{0}}$	&	3.508458$\times 10^{-7\phantom{0}}$	\\ \hline
3	&	1	&	6	&	7.014470$\times 10^{-9\phantom{0}}$	&	1.180811$\times 10^{-8\phantom{0}}$	&	1.866093$\times 10^{-7\phantom{0}}$	&	1.045302$\times 10^{-7\phantom{0}}$	\\ \hline
3	&	1	&	7	&	1.392344$\times 10^{-9\phantom{0}}$	&	2.201409$\times 10^{-9\phantom{0}}$	&	5.728150$\times 10^{-8\phantom{0}}$	&	2.993285$\times 10^{-8\phantom{0}}$	\\ \hline
4	&	1	&	0	&	-6.611595$\times 10^{-10}$	&	-3.354890$\times 10^{-9\phantom{0}}$	&	1.778200$\times 10^{-6\phantom{0}}$	&	-9.269538$\times 10^{-7\phantom{0}}$	\\ \hline
4	&	1	&	1	&	3.327272$\times 10^{-9\phantom{0}}$	&	1.078814$\times 10^{-8\phantom{0}}$	&	1.293272$\times 10^{-6\phantom{0}}$	&	-7.573044$\times 10^{-7\phantom{0}}$	\\ \hline
4	&	1	&	2	&	5.489708$\times 10^{-9\phantom{0}}$	&	1.347987$\times 10^{-8\phantom{0}}$	&	6.508828$\times 10^{-7\phantom{0}}$	&	-4.275336$\times 10^{-7\phantom{0}}$	\\ \hline
4	&	1	&	3	&	3.989895$\times 10^{-9\phantom{0}}$	&	8.004271$\times 10^{-9\phantom{0}}$	&	2.704943$\times 10^{-7\phantom{0}}$	&	-1.991900$\times 10^{-7\phantom{0}}$	\\ \hline
4	&	1	&	4	&	1.899103$\times 10^{-9\phantom{0}}$	&	3.250073$\times 10^{-9\phantom{0}}$	&	9.959920$\times 10^{-8\phantom{0}}$	&	-8.225436$\times 10^{-8\phantom{0}}$	\\ \hline
4	&	1	&	5	&	6.514413$\times 10^{-10}$	&	9.782216$\times 10^{-10}$	&	3.367305$\times 10^{-8\phantom{0}}$	&	-3.122972$\times 10^{-8\phantom{0}}$	\\ \hline
5	&	1	&	1	&	2.035370$\times 10^{-9\phantom{0}}$	&	6.185810$\times 10^{-9\phantom{0}}$	&	-1.963261$\times 10^{-8\phantom{0}}$	&	-1.023901$\times 10^{-7\phantom{0}}$	\\ \hline
5	&	1	&	3	&	-1.411206$\times 10^{-9\phantom{0}}$	&	-2.593480$\times 10^{-9\phantom{0}}$	&	-8.203217$\times 10^{-9\phantom{0}}$	&	-2.319941$\times 10^{-8\phantom{0}}$	\\ \hline
5	&	1	&	4	&	-1.152085$\times 10^{-9\phantom{0}}$	&	-1.783264$\times 10^{-9\phantom{0}}$	&	-3.940505$\times 10^{-9\phantom{0}}$	&	-8.863485$\times 10^{-9\phantom{0}}$	\\ \hline
    \end{tabular}
\end{table}

\begin{table}[htbp!]
    \centering
    \caption{List of partial amplitudes $C^{\pm}_{l2n}$ for the same orbit as in Table~\ref{tab:tab_m1}.}
    \label{tab:tab_m2}
    \begin{tabular}{|r|r|r|r|r|r|r|} \hline
\multicolumn{1}{|c}{$l$}	&	\multicolumn{1}{|c}{$m$}	&	\multicolumn{1}{|c}{$n$}	& \multicolumn{1}{|c}{${\rm Re}\{C^+_{lmn}\}$}	&	\multicolumn{1}{|c}{${\rm Im}\{C^+_{lmn}\}$}	&	\multicolumn{1}{|c}{${\rm Re}\{C^-_{lmn}\}$}	&	\multicolumn{1}{|c|}{${\rm Im}\{C^-_{lmn}\}$}	\\ \hline
2	&	2	&	-4	&	-1.646357$\times 10^{-9\phantom{0}}$	&	-2.613368$\times 10^{-10}$	&	2.350051$\times 10^{-8\phantom{0}}$	&	1.728856$\times 10^{-9\phantom{0}}$	\\ \hline
2	&	2	&	-3	&	-5.171809$\times 10^{-9\phantom{0}}$	&	-3.499305$\times 10^{-10}$	&	1.187803$\times 10^{-7\phantom{0}}$	&	1.441103$\times 10^{-8\phantom{0}}$	\\ \hline
2	&	2	&	-2	&	1.563190$\times 10^{-7\phantom{0}}$	&	-1.253110$\times 10^{-8\phantom{0}}$	&	2.380182$\times 10^{-6\phantom{0}}$	&	3.978260$\times 10^{-7\phantom{0}}$	\\ \hline
2	&	2	&	-1	&	-4.066035$\times 10^{-5\phantom{0}}$	&	6.801245$\times 10^{-6\phantom{0}}$	&	-9.742181$\times 10^{-5\phantom{0}}$	&	-2.076711$\times 10^{-5\phantom{0}}$	\\ \hline
2	&	2	&	0	&	3.858210$\times 10^{-4\phantom{0}}$	&	-8.824264$\times 10^{-5\phantom{0}}$	&	5.307355$\times 10^{-4\phantom{0}}$	&	1.379732$\times 10^{-4\phantom{0}}$	\\ \hline
2	&	2	&	1	&	3.970205$\times 10^{-4\phantom{0}}$	&	-1.089118$\times 10^{-4\phantom{0}}$	&	4.287891$\times 10^{-4\phantom{0}}$	&	1.314682$\times 10^{-4\phantom{0}}$	\\ \hline
2	&	2	&	2	&	2.406528$\times 10^{-4\phantom{0}}$	&	-7.404687$\times 10^{-5\phantom{0}}$	&	2.175548$\times 10^{-4\phantom{0}}$	&	7.698517$\times 10^{-5\phantom{0}}$	\\ \hline
2	&	2	&	3	&	1.138352$\times 10^{-4\phantom{0}}$	&	-3.764006$\times 10^{-5\phantom{0}}$	&	8.960040$\times 10^{-5\phantom{0}}$	&	3.602218$\times 10^{-5\phantom{0}}$	\\ \hline
2	&	2	&	4	&	4.644175$\times 10^{-5\phantom{0}}$	&	-1.599176$\times 10^{-5\phantom{0}}$	&	3.266409$\times 10^{-5\phantom{0}}$	&	1.474535$\times 10^{-5\phantom{0}}$	\\ \hline
2	&	2	&	5	&	1.715753$\times 10^{-5\phantom{0}}$	&	-5.997531$\times 10^{-6\phantom{0}}$	&	1.098058$\times 10^{-5\phantom{0}}$	&	5.516737$\times 10^{-6\phantom{0}}$	\\ \hline
2	&	2	&	6	&	5.901325$\times 10^{-6\phantom{0}}$	&	-2.047506$\times 10^{-6\phantom{0}}$	&	3.482534$\times 10^{-6\phantom{0}}$	&	1.934221$\times 10^{-6\phantom{0}}$	\\ \hline
2	&	2	&	7	&	1.922706$\times 10^{-6\phantom{0}}$	&	-6.480880$\times 10^{-7\phantom{0}}$	&	1.056831$\times 10^{-6\phantom{0}}$	&	6.456290$\times 10^{-7\phantom{0}}$	\\ \hline
2	&	2	&	8	&	6.002221$\times 10^{-7\phantom{0}}$	&	-1.923652$\times 10^{-7\phantom{0}}$	&	3.097499$\times 10^{-7\phantom{0}}$	&	2.073768$\times 10^{-7\phantom{0}}$	\\ \hline
2	&	2	&	9	&	1.810496$\times 10^{-7\phantom{0}}$	&	-5.403652$\times 10^{-8\phantom{0}}$	&	8.825836$\times 10^{-8\phantom{0}}$	&	6.461080$\times 10^{-8\phantom{0}}$	\\ \hline
2	&	2	&	10	&	5.321319$\times 10^{-8\phantom{0}}$	&	-1.460369$\times 10^{-8\phantom{0}}$	&	2.456752$\times 10^{-8\phantom{0}}$	&	1.967628$\times 10^{-8\phantom{0}}$	\\ \hline
2	&	2	&	11	&	1.545017$\times 10^{-8\phantom{0}}$	&	-3.880442$\times 10^{-9\phantom{0}}$	&	6.695307$\times 10^{-9\phantom{0}}$	&	5.913216$\times 10^{-9\phantom{0}}$	\\ \hline
2	&	2	&	12	&	4.635476$\times 10^{-9\phantom{0}}$	&	-9.455299$\times 10^{-10}$	&	1.752615$\times 10^{-9\phantom{0}}$	&	1.779985$\times 10^{-9\phantom{0}}$	\\ \hline
2	&	2	&	13	&	1.168229$\times 10^{-9\phantom{0}}$	&	-1.713718$\times 10^{-10}$	&	4.702090$\times 10^{-10}$	&	4.920492$\times 10^{-10}$	\\ \hline
3	&	2	&	-1	&	-2.396250$\times 10^{-7\phantom{0}}$	&	5.060581$\times 10^{-8\phantom{0}}$	&	1.890430$\times 10^{-6\phantom{0}}$	&	-9.847625$\times 10^{-7\phantom{0}}$	\\ \hline
3	&	2	&	0	&	3.649975$\times 10^{-6\phantom{0}}$	&	-1.111405$\times 10^{-6\phantom{0}}$	&	3.187469$\times 10^{-5\phantom{0}}$	&	-1.653879$\times 10^{-5\phantom{0}}$	\\ \hline
3	&	2	&	1	&	4.889631$\times 10^{-6\phantom{0}}$	&	-1.883465$\times 10^{-6\phantom{0}}$	&	3.015258$\times 10^{-5\phantom{0}}$	&	-1.566972$\times 10^{-5\phantom{0}}$	\\ \hline
3	&	2	&	2	&	3.536010$\times 10^{-6\phantom{0}}$	&	-1.615521$\times 10^{-6\phantom{0}}$	&	1.793053$\times 10^{-5\phantom{0}}$	&	-9.383274$\times 10^{-6\phantom{0}}$	\\ \hline
3	&	2	&	3	&	1.888741$\times 10^{-6\phantom{0}}$	&	-9.843758$\times 10^{-7\phantom{0}}$	&	8.495665$\times 10^{-6\phantom{0}}$	&	-4.500460$\times 10^{-6\phantom{0}}$	\\ \hline
3	&	2	&	4	&	8.370027$\times 10^{-7\phantom{0}}$	&	-4.845333$\times 10^{-7\phantom{0}}$	&	3.503785$\times 10^{-6\phantom{0}}$	&	-1.888329$\times 10^{-6\phantom{0}}$	\\ \hline
3	&	2	&	5	&	3.260656$\times 10^{-7\phantom{0}}$	&	-2.055320$\times 10^{-7\phantom{0}}$	&	1.314468$\times 10^{-6\phantom{0}}$	&	-7.241796$\times 10^{-7\phantom{0}}$	\\ \hline
3	&	2	&	6	&	1.154284$\times 10^{-7\phantom{0}}$	&	-7.798475$\times 10^{-8\phantom{0}}$	&	4.601592$\times 10^{-7\phantom{0}}$	&	-2.603261$\times 10^{-7\phantom{0}}$	\\ \hline
3	&	2	&	7	&	3.790759$\times 10^{-8\phantom{0}}$	&	-2.708873$\times 10^{-8\phantom{0}}$	&	1.527743$\times 10^{-7\phantom{0}}$	&	-8.912589$\times 10^{-8\phantom{0}}$	\\ \hline
3	&	2	&	8	&	1.170216$\times 10^{-8\phantom{0}}$	&	-8.743846$\times 10^{-9\phantom{0}}$	&	4.863519$\times 10^{-8\phantom{0}}$	&	-2.937305$\times 10^{-8\phantom{0}}$	\\ \hline
3	&	2	&	9	&	3.415775$\times 10^{-9\phantom{0}}$	&	-2.649605$\times 10^{-9\phantom{0}}$	&	1.496262$\times 10^{-8\phantom{0}}$	&	-9.389021$\times 10^{-9\phantom{0}}$	\\ \hline
3	&	2	&	10	&	9.324112$\times 10^{-10}$	&	-7.609938$\times 10^{-10}$	&	4.474358$\times 10^{-9\phantom{0}}$	&	-2.926757$\times 10^{-9\phantom{0}}$	\\ \hline
4	&	2	&	-1	&	-4.819515$\times 10^{-8\phantom{0}}$	&	1.102927$\times 10^{-8\phantom{0}}$	&	1.244599$\times 10^{-7\phantom{0}}$	&	-4.205354$\times 10^{-7\phantom{0}}$	\\ \hline
4	&	2	&	0	&	1.025155$\times 10^{-6\phantom{0}}$	&	-3.437785$\times 10^{-7\phantom{0}}$	&	1.114267$\times 10^{-6\phantom{0}}$	&	-4.056764$\times 10^{-6\phantom{0}}$	\\ \hline
4	&	2	&	1	&	3.046028$\times 10^{-7\phantom{0}}$	&	-1.313240$\times 10^{-7\phantom{0}}$	&	9.753818$\times 10^{-7\phantom{0}}$	&	-3.881701$\times 10^{-6\phantom{0}}$	\\ \hline
4	&	2	&	2	&	-1.647969$\times 10^{-7\phantom{0}}$	&	8.570891$\times 10^{-8\phantom{0}}$	&	5.323946$\times 10^{-7\phantom{0}}$	&	-2.357344$\times 10^{-6\phantom{0}}$	\\ \hline
4	&	2	&	3	&	-2.043076$\times 10^{-7\phantom{0}}$	&	1.234177$\times 10^{-7\phantom{0}}$	&	2.267391$\times 10^{-7\phantom{0}}$	&	-1.142665$\times 10^{-6\phantom{0}}$	\\ \hline
4	&	2	&	4	&	-1.208388$\times 10^{-7\phantom{0}}$	&	8.265904$\times 10^{-8\phantom{0}}$	&	8.153936$\times 10^{-8\phantom{0}}$	&	-4.822614$\times 10^{-7\phantom{0}}$	\\ \hline
4	&	2	&	5	&	-5.386171$\times 10^{-8\phantom{0}}$	&	4.095780$\times 10^{-8\phantom{0}}$	&	2.551314$\times 10^{-8\phantom{0}}$	&	-1.851227$\times 10^{-7\phantom{0}}$	\\ \hline
4	&	2	&	6	&	-2.029400$\times 10^{-8\phantom{0}}$	&	1.691104$\times 10^{-8\phantom{0}}$	&	6.938177$\times 10^{-9\phantom{0}}$	&	-6.628891$\times 10^{-8\phantom{0}}$	\\ \hline
4	&	2	&	7	&	-6.788401$\times 10^{-9\phantom{0}}$	&	6.130268$\times 10^{-9\phantom{0}}$	&	1.566483$\times 10^{-9\phantom{0}}$	&	-2.250158$\times 10^{-8\phantom{0}}$	\\ \hline
4	&	2	&	8	&	-2.068407$\times 10^{-9\phantom{0}}$	&	2.005378$\times 10^{-9\phantom{0}}$	&	2.389387$\times 10^{-10}$	&	-7.319933$\times 10^{-9\phantom{0}}$	\\ \hline
5	&	2	&	0	&	7.342146$\times 10^{-9\phantom{0}}$	&	-2.652971$\times 10^{-9\phantom{0}}$	&	-9.489133$\times 10^{-8\phantom{0}}$	&	-1.219765$\times 10^{-7\phantom{0}}$	\\ \hline
5	&	2	&	1	&	1.759299$\times 10^{-9\phantom{0}}$	&	-8.273581$\times 10^{-10}$	&	-1.010641$\times 10^{-7\phantom{0}}$	&	-1.194663$\times 10^{-7\phantom{0}}$	\\ \hline
5	&	2	&	2	&	-3.657312$\times 10^{-9\phantom{0}}$	&	2.102053$\times 10^{-9\phantom{0}}$	&	-7.026352$\times 10^{-8\phantom{0}}$	&	-7.612572$\times 10^{-8\phantom{0}}$	\\ \hline
5	&	2	&	3	&	-4.181570$\times 10^{-9\phantom{0}}$	&	2.831232$\times 10^{-9\phantom{0}}$	&	-3.875359$\times 10^{-8\phantom{0}}$	&	-3.833430$\times 10^{-8\phantom{0}}$	\\ \hline
5	&	2	&	4	&	-2.657863$\times 10^{-9\phantom{0}}$	&	2.069513$\times 10^{-9\phantom{0}}$	&	-1.841952$\times 10^{-8\phantom{0}}$	&	-1.655902$\times 10^{-8\phantom{0}}$	\\ \hline
5	&	2	&	5	&	-1.274114$\times 10^{-9\phantom{0}}$	&	1.121653$\times 10^{-9\phantom{0}}$	&	-7.882070$\times 10^{-9\phantom{0}}$	&	-6.404217$\times 10^{-9\phantom{0}}$	\\ \hline
    \end{tabular}
\end{table}

\begin{table}[htbp!]
    \centering
    \caption{List of partial amplitudes $C^{\pm}_{l3n}$ for the same orbit as in Table~\ref{tab:tab_m1}.}
    \label{tab:tab_m3}
    \begin{tabular}{|r|r|r|r|r|r|r|} \hline
\multicolumn{1}{|c}{$l$}	&	\multicolumn{1}{|c}{$m$}	&	\multicolumn{1}{|c}{$n$}	& \multicolumn{1}{|c}{${\rm Re}\{C^+_{lmn}\}$}	&	\multicolumn{1}{|c}{${\rm Im}\{C^+_{lmn}\}$}	&	\multicolumn{1}{|c}{${\rm Re}\{C^-_{lmn}\}$}	&	\multicolumn{1}{|c|}{${\rm Im}\{C^-_{lmn}\}$}	\\ \hline
3	&	3	&	-3	&	-1.004997$\times 10^{-9\phantom{0}}$	&	-7.760638$\times 10^{-9\phantom{0}}$	&	3.558851$\times 10^{-9\phantom{0}}$	&	1.588115$\times 10^{-8\phantom{0}}$	\\ \hline
3	&	3	&	-2	&	4.716853$\times 10^{-7\phantom{0}}$	&	2.003985$\times 10^{-6\phantom{0}}$	&	-3.882412$\times 10^{-7\phantom{0}}$	&	-1.922022$\times 10^{-6\phantom{0}}$	\\ \hline
3	&	3	&	-1	&	-1.300363$\times 10^{-5\phantom{0}}$	&	-4.030524$\times 10^{-5\phantom{0}}$	&	4.310309$\times 10^{-6\phantom{0}}$	&	2.366988$\times 10^{-5\phantom{0}}$	\\ \hline
3	&	3	&	0	&	5.305191$\times 10^{-5\phantom{0}}$	&	1.331426$\times 10^{-4\phantom{0}}$	&	-9.633994$\times 10^{-6\phantom{0}}$	&	-5.865711$\times 10^{-5\phantom{0}}$	\\ \hline
3	&	3	&	1	&	8.900651$\times 10^{-5\phantom{0}}$	&	1.911136$\times 10^{-4\phantom{0}}$	&	-1.135235$\times 10^{-5\phantom{0}}$	&	-7.650771$\times 10^{-5\phantom{0}}$	\\ \hline
3	&	3	&	2	&	7.612663$\times 10^{-5\phantom{0}}$	&	1.447861$\times 10^{-4\phantom{0}}$	&	-7.199807$\times 10^{-6\phantom{0}}$	&	-5.358218$\times 10^{-5\phantom{0}}$	\\ \hline
3	&	3	&	3	&	4.744130$\times 10^{-5\phantom{0}}$	&	8.189333$\times 10^{-5\phantom{0}}$	&	-3.482019$\times 10^{-6\phantom{0}}$	&	-2.851315$\times 10^{-5\phantom{0}}$	\\ \hline
3	&	3	&	4	&	2.434203$\times 10^{-5\phantom{0}}$	&	3.884905$\times 10^{-5\phantom{0}}$	&	-1.438478$\times 10^{-6\phantom{0}}$	&	-1.289482$\times 10^{-5\phantom{0}}$	\\ \hline
3	&	3	&	5	&	1.092213$\times 10^{-5\phantom{0}}$	&	1.635785$\times 10^{-5\phantom{0}}$	&	-5.363338$\times 10^{-7\phantom{0}}$	&	-5.227212$\times 10^{-6\phantom{0}}$	\\ \hline
3	&	3	&	6	&	4.438184$\times 10^{-6\phantom{0}}$	&	6.316315$\times 10^{-6\phantom{0}}$	&	-1.863735$\times 10^{-7\phantom{0}}$	&	-1.957612$\times 10^{-6\phantom{0}}$	\\ \hline
3	&	3	&	7	&	1.670164$\times 10^{-6\phantom{0}}$	&	2.283624$\times 10^{-6\phantom{0}}$	&	-6.164127$\times 10^{-8\phantom{0}}$	&	-6.902951$\times 10^{-7\phantom{0}}$	\\ \hline
3	&	3	&	8	&	5.909836$\times 10^{-7\phantom{0}}$	&	7.840535$\times 10^{-7\phantom{0}}$	&	-1.969217$\times 10^{-8\phantom{0}}$	&	-2.321446$\times 10^{-7\phantom{0}}$	\\ \hline
3	&	3	&	9	&	1.988825$\times 10^{-7\phantom{0}}$	&	2.581659$\times 10^{-7\phantom{0}}$	&	-6.141824$\times 10^{-9\phantom{0}}$	&	-7.513658$\times 10^{-8\phantom{0}}$	\\ \hline
3	&	3	&	10	&	6.409344$\times 10^{-8\phantom{0}}$	&	8.217920$\times 10^{-8\phantom{0}}$	&	-1.884840$\times 10^{-9\phantom{0}}$	&	-2.356231$\times 10^{-8\phantom{0}}$	\\ \hline
3	&	3	&	11	&	2.020696$\times 10^{-8\phantom{0}}$	&	2.531856$\times 10^{-8\phantom{0}}$	&	-5.716907$\times 10^{-10}$	&	-7.198380$\times 10^{-9\phantom{0}}$	\\ \hline
3	&	3	&	12	&	6.131692$\times 10^{-9\phantom{0}}$	&	7.692513$\times 10^{-9\phantom{0}}$	&	-1.717137$\times 10^{-10}$	&	-2.149742$\times 10^{-9\phantom{0}}$	\\ \hline
3	&	3	&	13	&	1.679868$\times 10^{-9\phantom{0}}$	&	2.245018$\times 10^{-9\phantom{0}}$	&	-5.246898$\times 10^{-11}$	&	-6.287407$\times 10^{-10}$	\\ \hline
4	&	3	&	-2	&	2.685489$\times 10^{-9\phantom{0}}$	&	1.028336$\times 10^{-8\phantom{0}}$	&	-2.488432$\times 10^{-8\phantom{0}}$	&	-2.373154$\times 10^{-8\phantom{0}}$	\\ \hline
4	&	3	&	-1	&	-1.147948$\times 10^{-7\phantom{0}}$	&	-3.145041$\times 10^{-7\phantom{0}}$	&	1.950153$\times 10^{-7\phantom{0}}$	&	1.857823$\times 10^{-7\phantom{0}}$	\\ \hline
4	&	3	&	0	&	6.392598$\times 10^{-7\phantom{0}}$	&	1.390373$\times 10^{-6\phantom{0}}$	&	-2.124982$\times 10^{-6\phantom{0}}$	&	-2.014665$\times 10^{-6\phantom{0}}$	\\ \hline
4	&	3	&	1	&	1.316270$\times 10^{-6\phantom{0}}$	&	2.399175$\times 10^{-6\phantom{0}}$	&	-2.891155$\times 10^{-6\phantom{0}}$	&	-2.717792$\times 10^{-6\phantom{0}}$	\\ \hline
4	&	3	&	2	&	1.323400$\times 10^{-6\phantom{0}}$	&	2.090102$\times 10^{-6\phantom{0}}$	&	-2.242899$\times 10^{-6\phantom{0}}$	&	-2.082785$\times 10^{-6\phantom{0}}$	\\ \hline
4	&	3	&	3	&	9.395536$\times 10^{-7\phantom{0}}$	&	1.315492$\times 10^{-6\phantom{0}}$	&	-1.324312$\times 10^{-6\phantom{0}}$	&	-1.210354$\times 10^{-6\phantom{0}}$	\\ \hline
4	&	3	&	4	&	5.364249$\times 10^{-7\phantom{0}}$	&	6.771760$\times 10^{-7\phantom{0}}$	&	-6.609932$\times 10^{-7\phantom{0}}$	&	-5.923846$\times 10^{-7\phantom{0}}$	\\ \hline
4	&	3	&	5	&	2.629183$\times 10^{-7\phantom{0}}$	&	3.032417$\times 10^{-7\phantom{0}}$	&	-2.939162$\times 10^{-7\phantom{0}}$	&	-2.573399$\times 10^{-7\phantom{0}}$	\\ \hline
4	&	3	&	6	&	1.149548$\times 10^{-7\phantom{0}}$	&	1.224621$\times 10^{-7\phantom{0}}$	&	-1.200513$\times 10^{-7\phantom{0}}$	&	-1.023073$\times 10^{-7\phantom{0}}$	\\ \hline
4	&	3	&	7	&	4.595850$\times 10^{-8\phantom{0}}$	&	4.564687$\times 10^{-8\phantom{0}}$	&	-4.593515$\times 10^{-8\phantom{0}}$	&	-3.795772$\times 10^{-8\phantom{0}}$	\\ \hline
4	&	3	&	8	&	1.708722$\times 10^{-8\phantom{0}}$	&	1.595567$\times 10^{-8\phantom{0}}$	&	-1.668689$\times 10^{-8\phantom{0}}$	&	-1.331898$\times 10^{-8\phantom{0}}$	\\ \hline
4	&	3	&	9	&	5.977998$\times 10^{-9\phantom{0}}$	&	5.284213$\times 10^{-9\phantom{0}}$	&	-5.810556$\times 10^{-9\phantom{0}}$	&	-4.462024$\times 10^{-9\phantom{0}}$	\\ \hline
4	&	3	&	10	&	1.986521$\times 10^{-9\phantom{0}}$	&	1.675756$\times 10^{-9\phantom{0}}$	&	-1.953240$\times 10^{-9\phantom{0}}$	&	-1.437145$\times 10^{-9\phantom{0}}$	\\ \hline
5	&	3	&	-2	&	7.907117$\times 10^{-10}$	&	2.853014$\times 10^{-9\phantom{0}}$	&	-6.478361$\times 10^{-9\phantom{0}}$	&	-3.587094$\times 10^{-10}$	\\ \hline
5	&	3	&	-1	&	-4.398249$\times 10^{-8\phantom{0}}$	&	-1.123248$\times 10^{-7\phantom{0}}$	&	1.317613$\times 10^{-8\phantom{0}}$	&	5.278265$\times 10^{-10}$	\\ \hline
5	&	3	&	0	&	3.218695$\times 10^{-7\phantom{0}}$	&	6.453677$\times 10^{-7\phantom{0}}$	&	-3.389273$\times 10^{-7\phantom{0}}$	&	-7.785256$\times 10^{-9\phantom{0}}$	\\ \hline
5	&	3	&	1	&	3.455801$\times 10^{-7\phantom{0}}$	&	5.738309$\times 10^{-7\phantom{0}}$	&	-4.536955$\times 10^{-7\phantom{0}}$	&	-1.859315$\times 10^{-9\phantom{0}}$	\\ \hline
5	&	3	&	2	&	1.711208$\times 10^{-7\phantom{0}}$	&	2.430505$\times 10^{-7\phantom{0}}$	&	-3.512562$\times 10^{-7\phantom{0}}$	&	5.820222$\times 10^{-9\phantom{0}}$	\\ \hline
5	&	3	&	3	&	3.896109$\times 10^{-8\phantom{0}}$	&	4.837181$\times 10^{-8\phantom{0}}$	&	-2.071569$\times 10^{-7\phantom{0}}$	&	8.090550$\times 10^{-9\phantom{0}}$	\\ \hline
5	&	3	&	4	&	-1.094717$\times 10^{-8\phantom{0}}$	&	-1.206677$\times 10^{-8\phantom{0}}$	&	-1.031213$\times 10^{-7\phantom{0}}$	&	6.536982$\times 10^{-9\phantom{0}}$	\\ \hline
5	&	3	&	5	&	-1.726012$\times 10^{-8\phantom{0}}$	&	-1.709111$\times 10^{-8\phantom{0}}$	&	-4.563110$\times 10^{-8\phantom{0}}$	&	4.089656$\times 10^{-9\phantom{0}}$	\\ \hline
5	&	3	&	6	&	-1.143495$\times 10^{-8\phantom{0}}$	&	-1.026788$\times 10^{-8\phantom{0}}$	&	-1.850203$\times 10^{-8\phantom{0}}$	&	2.179864$\times 10^{-9\phantom{0}}$	\\ \hline
5	&	3	&	7	&	-5.747543$\times 10^{-9\phantom{0}}$	&	-4.714943$\times 10^{-9\phantom{0}}$	&	-7.009564$\times 10^{-9\phantom{0}}$	&	1.037791$\times 10^{-9\phantom{0}}$	\\ \hline
5	&	3	&	8	&	-2.468744$\times 10^{-9\phantom{0}}$	&	-1.860621$\times 10^{-9\phantom{0}}$	&	-2.514576$\times 10^{-9\phantom{0}}$	&	4.537261$\times 10^{-10}$	\\ \hline
5	&	3	&	9	&	-9.509418$\times 10^{-10}$	&	-6.710533$\times 10^{-10}$	&	-8.623420$\times 10^{-10}$	&	1.854954$\times 10^{-10}$	\\ \hline
6	&	3	&	0	&	2.515496$\times 10^{-9\phantom{0}}$	&	4.721206$\times 10^{-9\phantom{0}}$	&	-8.984566$\times 10^{-9\phantom{0}}$	&	9.229790$\times 10^{-9\phantom{0}}$	\\ \hline
6	&	3	&	1	&	3.001379$\times 10^{-9\phantom{0}}$	&	4.619049$\times 10^{-9\phantom{0}}$	&	-1.151088$\times 10^{-8\phantom{0}}$	&	1.262076$\times 10^{-8\phantom{0}}$	\\ \hline
6	&	3	&	2	&	1.275195$\times 10^{-9\phantom{0}}$	&	1.660290$\times 10^{-9\phantom{0}}$	&	-9.081402$\times 10^{-9\phantom{0}}$	&	1.067007$\times 10^{-8\phantom{0}}$	\\ \hline
6	&	3	&	4	&	-7.270147$\times 10^{-10}$	&	-7.155788$\times 10^{-10}$	&	-2.769497$\times 10^{-9\phantom{0}}$	&	3.789745$\times 10^{-9\phantom{0}}$	\\ \hline
    \end{tabular}
\end{table}

\begin{table}[htbp!]
    \centering
    \caption{List of partial amplitudes $C^{\pm}_{l4n}$ for the same orbit as in Table~\ref{tab:tab_m1}.}
    \label{tab:tab_m4}
    \begin{tabular}{|r|r|r|r|r|r|r|} \hline
\multicolumn{1}{|c}{$l$}	&	\multicolumn{1}{|c}{$m$}	&	\multicolumn{1}{|c}{$n$}	& \multicolumn{1}{|c}{${\rm Re}\{C^+_{lmn}\}$}	&	\multicolumn{1}{|c}{${\rm Im}\{C^+_{lmn}\}$}	&	\multicolumn{1}{|c}{${\rm Re}\{C^-_{lmn}\}$}	&	\multicolumn{1}{|c|}{${\rm Im}\{C^-_{lmn}\}$}	\\ \hline
4	&	4	&	-3	&	1.030232$\times 10^{-7\phantom{0}}$	&	-2.987138$\times 10^{-8\phantom{0}}$	&	2.594938$\times 10^{-8\phantom{0}}$	&	-1.364431$\times 10^{-8\phantom{0}}$	\\ \hline
4	&	4	&	-2	&	-3.167459$\times 10^{-6\phantom{0}}$	&	1.235891$\times 10^{-6\phantom{0}}$	&	-5.564356$\times 10^{-7\phantom{0}}$	&	2.832212$\times 10^{-7\phantom{0}}$	\\ \hline
4	&	4	&	-1	&	2.493681$\times 10^{-5\phantom{0}}$	&	-1.203041$\times 10^{-5\phantom{0}}$	&	3.475072$\times 10^{-6\phantom{0}}$	&	-1.718247$\times 10^{-6\phantom{0}}$	\\ \hline
4	&	4	&	0	&	-3.730632$\times 10^{-5\phantom{0}}$	&	2.123841$\times 10^{-5\phantom{0}}$	&	-4.314063$\times 10^{-6\phantom{0}}$	&	2.079650$\times 10^{-6\phantom{0}}$	\\ \hline
4	&	4	&	1	&	-8.092430$\times 10^{-5\phantom{0}}$	&	5.277164$\times 10^{-5\phantom{0}}$	&	-9.355450$\times 10^{-6\phantom{0}}$	&	4.413471$\times 10^{-6\phantom{0}}$	\\ \hline
4	&	4	&	2	&	-7.619630$\times 10^{-5\phantom{0}}$	&	5.574275$\times 10^{-5\phantom{0}}$	&	-8.650047$\times 10^{-6\phantom{0}}$	&	4.008745$\times 10^{-6\phantom{0}}$	\\ \hline
4	&	4	&	3	&	-5.089773$\times 10^{-5\phantom{0}}$	&	4.112353$\times 10^{-5\phantom{0}}$	&	-5.704990$\times 10^{-6\phantom{0}}$	&	2.607445$\times 10^{-6\phantom{0}}$	\\ \hline
4	&	4	&	4	&	-2.773635$\times 10^{-5\phantom{0}}$	&	2.444478$\times 10^{-5\phantom{0}}$	&	-3.087320$\times 10^{-6\phantom{0}}$	&	1.397110$\times 10^{-6\phantom{0}}$	\\ \hline
4	&	4	&	5	&	-1.316378$\times 10^{-5\phantom{0}}$	&	1.252512$\times 10^{-5\phantom{0}}$	&	-1.461746$\times 10^{-6\phantom{0}}$	&	6.575647$\times 10^{-7\phantom{0}}$	\\ \hline
4	&	4	&	6	&	-5.647802$\times 10^{-6\phantom{0}}$	&	5.750074$\times 10^{-6\phantom{0}}$	&	-6.277339$\times 10^{-7\phantom{0}}$	&	2.818301$\times 10^{-7\phantom{0}}$	\\ \hline
4	&	4	&	7	&	-2.243064$\times 10^{-6\phantom{0}}$	&	2.424266$\times 10^{-6\phantom{0}}$	&	-2.501089$\times 10^{-7\phantom{0}}$	&	1.125120$\times 10^{-7\phantom{0}}$	\\ \hline
4	&	4	&	8	&	-8.381942$\times 10^{-7\phantom{0}}$	&	9.546192$\times 10^{-7\phantom{0}}$	&	-9.388208$\times 10^{-8\phantom{0}}$	&	4.248119$\times 10^{-8\phantom{0}}$	\\ \hline
4	&	4	&	9	&	-2.980940$\times 10^{-7\phantom{0}}$	&	3.554223$\times 10^{-7\phantom{0}}$	&	-3.356380$\times 10^{-8\phantom{0}}$	&	1.533473$\times 10^{-8\phantom{0}}$	\\ \hline
4	&	4	&	10	&	-1.020168$\times 10^{-7\phantom{0}}$	&	1.261960$\times 10^{-7\phantom{0}}$	&	-1.152115$\times 10^{-8\phantom{0}}$	&	5.334443$\times 10^{-9\phantom{0}}$	\\ \hline
4	&	4	&	11	&	-3.378335$\times 10^{-8\phantom{0}}$	&	4.299271$\times 10^{-8\phantom{0}}$	&	-3.820558$\times 10^{-9\phantom{0}}$	&	1.799074$\times 10^{-9\phantom{0}}$	\\ \hline
4	&	4	&	12	&	-1.082123$\times 10^{-8\phantom{0}}$	&	1.416903$\times 10^{-8\phantom{0}}$	&	-1.229874$\times 10^{-9\phantom{0}}$	&	5.909993$\times 10^{-10}$	\\ \hline
4	&	4	&	13	&	-2.922436$\times 10^{-9\phantom{0}}$	&	4.787284$\times 10^{-9\phantom{0}}$	&	-3.860830$\times 10^{-10}$	&	1.899575$\times 10^{-10}$	\\ \hline
4	&	4	&	14	&	-1.034014$\times 10^{-9\phantom{0}}$	&	1.406801$\times 10^{-9\phantom{0}}$	&	-1.182908$\times 10^{-10}$	&	5.976449$\times 10^{-11}$	\\ \hline
5	&	4	&	-2	&	-1.875445$\times 10^{-8\phantom{0}}$	&	7.933698$\times 10^{-9\phantom{0}}$	&	-3.023775$\times 10^{-9\phantom{0}}$	&	5.361340$\times 10^{-9\phantom{0}}$	\\ \hline
5	&	4	&	-1	&	1.928140$\times 10^{-7\phantom{0}}$	&	-1.021033$\times 10^{-7\phantom{0}}$	&	3.348964$\times 10^{-8\phantom{0}}$	&	-5.960966$\times 10^{-8\phantom{0}}$	\\ \hline
5	&	4	&	0	&	-3.581359$\times 10^{-7\phantom{0}}$	&	2.267803$\times 10^{-7\phantom{0}}$	&	-1.071155$\times 10^{-7\phantom{0}}$	&	1.920896$\times 10^{-7\phantom{0}}$	\\ \hline
5	&	4	&	1	&	-8.885218$\times 10^{-7\phantom{0}}$	&	6.538335$\times 10^{-7\phantom{0}}$	&	-2.097613$\times 10^{-7\phantom{0}}$	&	3.803623$\times 10^{-7\phantom{0}}$	\\ \hline
5	&	4	&	2	&	-9.357029$\times 10^{-7\phantom{0}}$	&	7.847094$\times 10^{-7\phantom{0}}$	&	-2.025645$\times 10^{-7\phantom{0}}$	&	3.727819$\times 10^{-7\phantom{0}}$	\\ \hline
5	&	4	&	3	&	-6.840485$\times 10^{-7\phantom{0}}$	&	6.445750$\times 10^{-7\phantom{0}}$	&	-1.413811$\times 10^{-7\phantom{0}}$	&	2.650546$\times 10^{-7\phantom{0}}$	\\ \hline
5	&	4	&	4	&	-4.007050$\times 10^{-7\phantom{0}}$	&	4.196831$\times 10^{-7\phantom{0}}$	&	-8.077376$\times 10^{-8\phantom{0}}$	&	1.548634$\times 10^{-7\phantom{0}}$	\\ \hline
5	&	4	&	5	&	-2.013663$\times 10^{-7\phantom{0}}$	&	2.324005$\times 10^{-7\phantom{0}}$	&	-4.013179$\times 10^{-8\phantom{0}}$	&	7.900370$\times 10^{-8\phantom{0}}$	\\ \hline
5	&	4	&	6	&	-9.029619$\times 10^{-8\phantom{0}}$	&	1.140120$\times 10^{-7\phantom{0}}$	&	-1.795737$\times 10^{-8\phantom{0}}$	&	3.645112$\times 10^{-8\phantom{0}}$	\\ \hline
5	&	4	&	7	&	-3.705473$\times 10^{-8\phantom{0}}$	&	5.087027$\times 10^{-8\phantom{0}}$	&	-7.399921$\times 10^{-9\phantom{0}}$	&	1.555748$\times 10^{-8\phantom{0}}$	\\ \hline
5	&	4	&	8	&	-1.416211$\times 10^{-8\phantom{0}}$	&	2.102079$\times 10^{-8\phantom{0}}$	&	-2.851190$\times 10^{-9\phantom{0}}$	&	6.238100$\times 10^{-9\phantom{0}}$	\\ \hline
5	&	4	&	9	&	-5.102297$\times 10^{-9\phantom{0}}$	&	8.148365$\times 10^{-9\phantom{0}}$	&	-1.038250$\times 10^{-9\phantom{0}}$	&	2.376195$\times 10^{-9\phantom{0}}$	\\ \hline
5	&	4	&	10	&	-1.757070$\times 10^{-9\phantom{0}}$	&	2.997933$\times 10^{-9\phantom{0}}$	&	-3.601133$\times 10^{-10}$	&	8.670269$\times 10^{-10}$	\\ \hline
5	&	4	&	11	&	-5.830642$\times 10^{-10}$	&	1.054002$\times 10^{-9\phantom{0}}$	&	-1.196530$\times 10^{-10}$	&	3.049759$\times 10^{-10}$	\\ \hline
6	&	4	&	-2	&	-8.666245$\times 10^{-9\phantom{0}}$	&	3.877093$\times 10^{-9\phantom{0}}$	&	1.065444$\times 10^{-10}$	&	7.051137$\times 10^{-10}$	\\ \hline
6	&	4	&	-1	&	1.052404$\times 10^{-7\phantom{0}}$	&	-5.945050$\times 10^{-8\phantom{0}}$	&	-9.799602$\times 10^{-10}$	&	-5.966157$\times 10^{-9\phantom{0}}$	\\ \hline
6	&	4	&	0	&	-2.657233$\times 10^{-7\phantom{0}}$	&	1.812153$\times 10^{-7\phantom{0}}$	&	4.508872$\times 10^{-9\phantom{0}}$	&	2.519233$\times 10^{-8\phantom{0}}$	\\ \hline
6	&	4	&	1	&	-3.867479$\times 10^{-7\phantom{0}}$	&	3.097707$\times 10^{-7\phantom{0}}$	&	9.373712$\times 10^{-9\phantom{0}}$	&	4.799473$\times 10^{-8\phantom{0}}$	\\ \hline
6	&	4	&	2	&	-2.693825$\times 10^{-7\phantom{0}}$	&	2.488393$\times 10^{-7\phantom{0}}$	&	9.932362$\times 10^{-9\phantom{0}}$	&	4.656984$\times 10^{-8\phantom{0}}$	\\ \hline
6	&	4	&	3	&	-1.291963$\times 10^{-7\phantom{0}}$	&	1.359028$\times 10^{-7\phantom{0}}$	&	7.665334$\times 10^{-9\phantom{0}}$	&	3.290682$\times 10^{-8\phantom{0}}$	\\ \hline
6	&	4	&	4	&	-4.680907$\times 10^{-8\phantom{0}}$	&	5.556090$\times 10^{-8\phantom{0}}$	&	4.860330$\times 10^{-9\phantom{0}}$	&	1.910878$\times 10^{-8\phantom{0}}$	\\ \hline
6	&	4	&	5	&	-1.257609$\times 10^{-8\phantom{0}}$	&	1.673105$\times 10^{-8\phantom{0}}$	&	2.686650$\times 10^{-9\phantom{0}}$	&	9.678828$\times 10^{-9\phantom{0}}$	\\ \hline
6	&	4	&	6	&	-1.863538$\times 10^{-9\phantom{0}}$	&	2.764762$\times 10^{-9\phantom{0}}$	&	1.340226$\times 10^{-9\phantom{0}}$	&	4.427345$\times 10^{-9\phantom{0}}$	\\ \hline
6	&	4	&	7	&	4.381872$\times 10^{-10}$	&	-7.223455$\times 10^{-10}$	&	6.169229$\times 10^{-10}$	&	1.870247$\times 10^{-9\phantom{0}}$	\\ \hline
6	&	4	&	8	&	5.248114$\times 10^{-10}$	&	-9.583384$\times 10^{-10}$	&	2.660845$\times 10^{-10}$	&	7.408825$\times 10^{-10}$	\\ \hline
7	&	4	&	0	&	-1.681887$\times 10^{-9\phantom{0}}$	&	1.221218$\times 10^{-9\phantom{0}}$	&	8.419294$\times 10^{-10}$	&	6.223333$\times 10^{-10}$	\\ \hline
7	&	4	&	1	&	-2.629027$\times 10^{-9\phantom{0}}$	&	2.263383$\times 10^{-9\phantom{0}}$	&	1.494752$\times 10^{-9\phantom{0}}$	&	1.045267$\times 10^{-9\phantom{0}}$	\\ \hline
7	&	4	&	2	&	-1.825947$\times 10^{-9\phantom{0}}$	&	1.832837$\times 10^{-9\phantom{0}}$	&	1.524851$\times 10^{-9\phantom{0}}$	&	1.004491$\times 10^{-9\phantom{0}}$	\\ \hline
7	&	4	&	3	&	-7.525698$\times 10^{-10}$	&	8.711420$\times 10^{-10}$	&	1.156468$\times 10^{-9\phantom{0}}$	&	7.142646$\times 10^{-10}$	\\ \hline
    \end{tabular}
\end{table}

\twocolumngrid

\bibliographystyle{unsrt}
\bibliography{paper}

\end{document}